# The Chandra Source Catalog Normal Galaxy Sample

Dong-Woo Kim[1], Alyssa Cassity[2], Binod Bhatt[3],
Giuseppina Fabbiano[1], Juan Rafael Martinez Galarza[1], Ewan O'Sullivan[1], Arnold Rots[1]

1. Center for Astrophysics | Harvard and Smithsonian
2. University of British Columbia
3. University of Rochester

(June 14, 2023)

### Abstract

We present an extensive and well-characterized Chandra X-ray Galaxy Catalog (CGC) of 8547 galaxy candidates in the redshift range z ~ 0.04 - 0.7, optical luminosity $10^{10}$ - $10^{11}$ $Lr_\odot$, and X-ray luminosity (0.5-7 keV) $L_X$ = $2 \times 10^{40}$ - $2 \times 10^{43}$ erg s$^{-1}$. We estimate ~5% false match fraction and contamination by QSOs. The CGC was extracted from the Chandra Source Catalog version 2 (CSC2) by cross-correlating with optical and IR all-sky survey data, including SDSS, PanSTARRS, DESI Legacy, and WISE. Our selection makes use of two main criteria that we have tested on the subsample with optical spectroscopical identification. (1) A joint selection based on X-ray luminosity ($L_X$) and X-ray to optical flux ratio ($F_{XO}$), which recovers 63% of the spectroscopically classified galaxies with a small contamination fraction (7%), a significant improvement over methods using $L_X$ or $F_{XO}$ alone (< 50% recovery). (2) A joint W1-W2 ($W_{12}$) WISE color and $L_X$ selection that proves effective in excluding QSOs and improves our selection by recovering 72% of the spectroscopically classified galaxies and reducing the contamination fraction (4%). Of the CGC, 24% was selected by means of optical spectroscopy; 30% on the basis of $L_X$, $F_{XO}$, and $W_{12}$; and 46% by using either the $L_X$-$F_{XO}$ or the $L_X$-$W_{12}$ selection criteria. We have individually examined the data for galaxies with z < 0.1, which may include more than one CSC2 X-ray source, leading to the exclusion of 110 local galaxies. Our catalog also includes near-IR and UV data and galaxy morphological types.

Unified Astronomy Thesaurus concepts: Galaxies (573); X-ray astronomy (1810); X-ray sources (1822); Classification (1907); X-ray surveys (1824); Active galactic nuclei (16); Catalogs (205)

# 1. INTRODUCTION

Over the past 20 years, Chandra X-ray observations have led to a new understanding in several areas of astronomy and astrophysics (see Wilkes & Tucker 2019). The unprecedented high spatial resolution of Chandra[1] provides the ability to resolve fine structures at a sub-arcsec scale and detect the faintest X-ray sources. These capabilities have taken to a new level the study of galaxies and their components: the hot ISM, populations of low and high mass X-ray binaries (XRB), and nuclear activity. These Chandra results have highlighted how the X-ray emission is important for our understanding of the formation and evolution of galaxies (e.g., Kim and Pellegrini 2012; Fabbiano 2019). For example, the hot ISM, which was previously considered as smooth and featureless, has been shown to have complex X-ray morphologies and spectral properties, pointing to fundamental evolutionary mechanisms, such as AGN feedback, merging history, accretion, stripping, and star formation (SF) and its quenching (Nardini, Kim & Pellegrini 2022; Fabbiano & Elvis 2022). Similarly, the evolution of XRBs, as well as AGNs, may also result in substantial energy input (feedback) into the ISM (Fabian 2012; Choi et al. 2017; Ciotti et al. 2017).

The aim of this work is to build an extensive, unbiased catalog of X-ray-selected normal (not AGN-dominated) galaxies, the CGC. The CGC is needed to put onto a stronger statistical footing the scaling relations between observed parameters (e.g., $L_X$, T, multi-wavelength luminosity and colors) and fundamental physical parameters of galaxies (e.g., Mass, SF) that have been discussed in the literature (Boroson, Kim & Fabbiano 2011; Li & Wang 2013; Kim & Fabbiano 2015; Bogdan & Goulding 2015) and also to produce a highly significant X-ray luminosity function of galaxies. While the Chandra observations of nearby galaxies have been instrumental in producing the luminosity function of XRB populations and thereby exploring the evolution of these binary stellar systems (e.g., Kim et al. 2009; Kim & Fabbiano 2010; Mineo et al. 2012; Lehmer et al. 2016), it has not been possible to fully explore the X-ray luminosities of different types of galaxies, given the limited sample size available (e.g., Kim et al. 2006). Numerical simulations can broadly reproduce the scaling relations of hot gas properties (e.g., $L_{X,GAS}$ and $T_{GAS}$), emphasizing the importance of stellar and AGN feedback in galaxy formation and evolution (Choi et al. 2017; Ciotti et al. 2017; Vogelsberger et al. 2020; Kelly et al. 2021). The CGC provides the means to pose significantly stricter constraints on the observed parameter space.

The Chandra Multiwavelength Project (ChaMP) conducted the first serendipitous survey of the sky with Chandra, producing ~7000 sources from the sky area of 10 deg$^2$ (Kim et al. 2007), which were identified with stars, galaxies, AGNs, and galaxy clusters. A minority of these sources, 136, were identified with normal galaxies (Kim et al. 2006), a result consistent with that of the COSMOS X-ray survey, where normal galaxies comprise about a few percent of the sample (Marchesi et al. 2016). Therefore, to increase the normal galaxy sample significantly, we need to survey a much larger region of the sky. The Chandra Source Catalog Version 2 (CSC2) provides this opportunity. The CSC2 includes 317,167 unique X-ray sources with accurate positions and high-quality photometric, spectral, and temporal properties (Evans et al. 2010). To extract an expanded sample of normal galaxies, we have crossmatched the CSC2 with a suite of optical/IR/UV catalogs. Our selection makes use of accurate positions and selection criteria aimed at minimizing sample contamination by AGNs/QSOs.

This paper is organized as follows. In Section 2, we describe the X-ray source selection. In Section 3, we describe the optical/IR/UV catalogs which we use to crossmatch X-ray sources and classify galaxies. In Section 4, we present our cross-matching procedures and the match statistics, including false match rates. In Section 5, we describe the galaxy-finding algorithms and the galaxy properties in the multi-wavelength parameter space. In Section 6, we present our final Chandra X-ray Galaxy Catalog (CGC). In Section 7, we list the science topics which we plan to explore with our CGC.

---

[1] https://asc.harvard.edu/proposer/POG/

Throughout the paper, we adopt the following cosmological parameters: $H_o$= 69.6 (km/s)/Mpc, $\Omega_M$= 0.286, and $\Omega_\Lambda$= 0.714.

## 2. THE CHANDRA X-RAY SOURCE CATALOG

The main advantage of the Chandra Observatory is its unprecedented high spatial resolution – the capability to resolve individual sources in a crowded field and to determine their positions on a sub-arcsec scale. This resolution also results in the best detection efficiency for the faintest X-ray sources in the sky. The CSC2[2], released in 2019, makes use of the first 15 years of Chandra observations (1999 – 2014), co-added in overlapping regions of the sky to achieve the best sensitivity. The CSC2 includes 317,167 unique X-ray sources with accurate positions and high-quality photometric, spectral, and temporal properties. Because of the angular resolution of Chandra, the CSC2 provides the most precise source position and is the least affected by source confusion among the X-ray catalogs. Figure 1 shows the CSC2 sources plotted in galactic coordinates on the Aitoff projection.

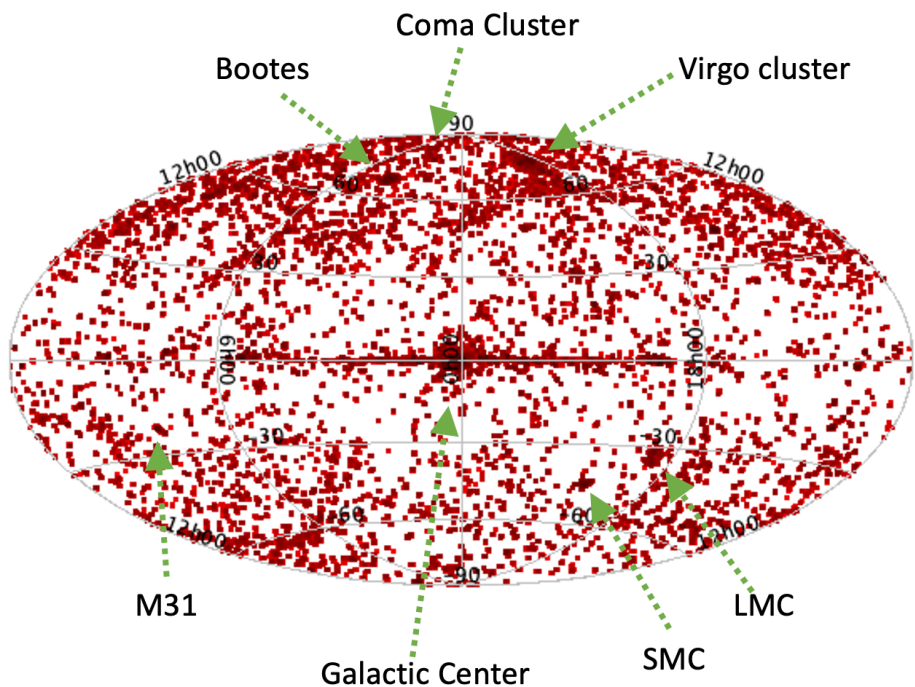

Fig. 1. X-ray sources in CSC2 are plotted in galactic coordinates (l, b). Also marked are a few well-known objects.

---

[2] https://cxc.harvard.edu/csc/

Out of the 317,167 CSC2 sources, we selected only those with high data quality by applying the following additional filters:
- Streak flag = False (to exclude sources located on an ACIS readout streak)
- Pile-up flag = False (to exclude sources with ACIS pile-up fraction > 10%)
- Saturation source flag = False (to exclude sources for which all stacked observation detections are significantly piled-up)

Moreover, to minimize the false-association rate to ~5%, we excluded crowded sky regions ($|b| <$ 15°, and higher galactic latitude crowded fields, typically including Local Group galaxies or galactic star clusters - see Section 4). We excluded 1299 highly extended (> 30 arcsec) sources, called Convex-Hull sources[3]. Their source properties are considered as an alpha-release in CSC2 and are not as reliable as the compact sources. We also excluded the CSC2 sources for which the positions of the optical/IR counterparts differed by more than 1 or 2 arcseconds (see Section 4 for details). This screening resulted in a catalog of 117,358 high quality, high-galactic-latitude 'point-like' or moderately extended sources.

## 3. MULTI-WAVELENGTH CATALOGS

We crossmatched the X-ray sources with optical/IR/UV catalogs to identify galaxies and obtain multi-wavelength data using the all sky surveys listed below. To illustrate the sky coverage of each catalog, in Figure 2 we show the matched sources from each catalog in galactic coordinates on the Aitoff projection.

### 3.1 SDSS[4]

The spectroscopic and photometric catalogs in DR16 include 5 million (M) and 800M sources in the SpecObj and PhotoObj tables, respectively. Among the spectroscopic sample, ~900K galaxies have additional information on their morphological types from the Galaxy Zoo[5]. Those in the PhotoObj table (measured in 5 bands - u, g, r, i, and z) consist of 430M point-like objects (type=6 or star) and 370M extended objects (type=3 or galaxy). Among the photometric sample, the photometric redshifts of ~200M objects are available (Beck et al. 2016)

### 3.2 PanSTARRS[6]

The DR2 includes 3 billion (B) sources from the 3π sky with δ > -30°, with optical photometric data in 5 bands (g, r, i, z, y) and surface brightness fits to morphological models (e.g., exponential, de Vaucouleurs). Although their spectroscopic redshifts are unavailable, photometric redshifts are available for 2.9B sources (Beck et al. 2021). Additionally, 1.6M galaxies have broad morphological types by Goddard et al. (2020).

### 3.3 DESI Legacy[7]

---

[3] https://cxc.harvard.edu/csc/columns/chs_properties.html
[4] https://www.sdss.org/dr16/
[5] https://www.zooniverse.org/projects/zookeeper/galaxy-zoo/
[6] https://panstarrs.stsci.edu
[7] https://www.legacysurvey.org

The DR8 catalog includes 2 B sources from 14,000 deg² of the extragalactic sky in -18° < δ < +84° and |b| > 18°, including the southern sky, which SDSS does not cover (see Fig. 2). This catalog has optical photometric data in 3 bands (g, r, z). For 0.17 B sources, the photometric redshifts and fitting results of various morphological models (e.g., PSF, exponential, de Vaucouleurs, and composite) are available (Zou et al. 2019).

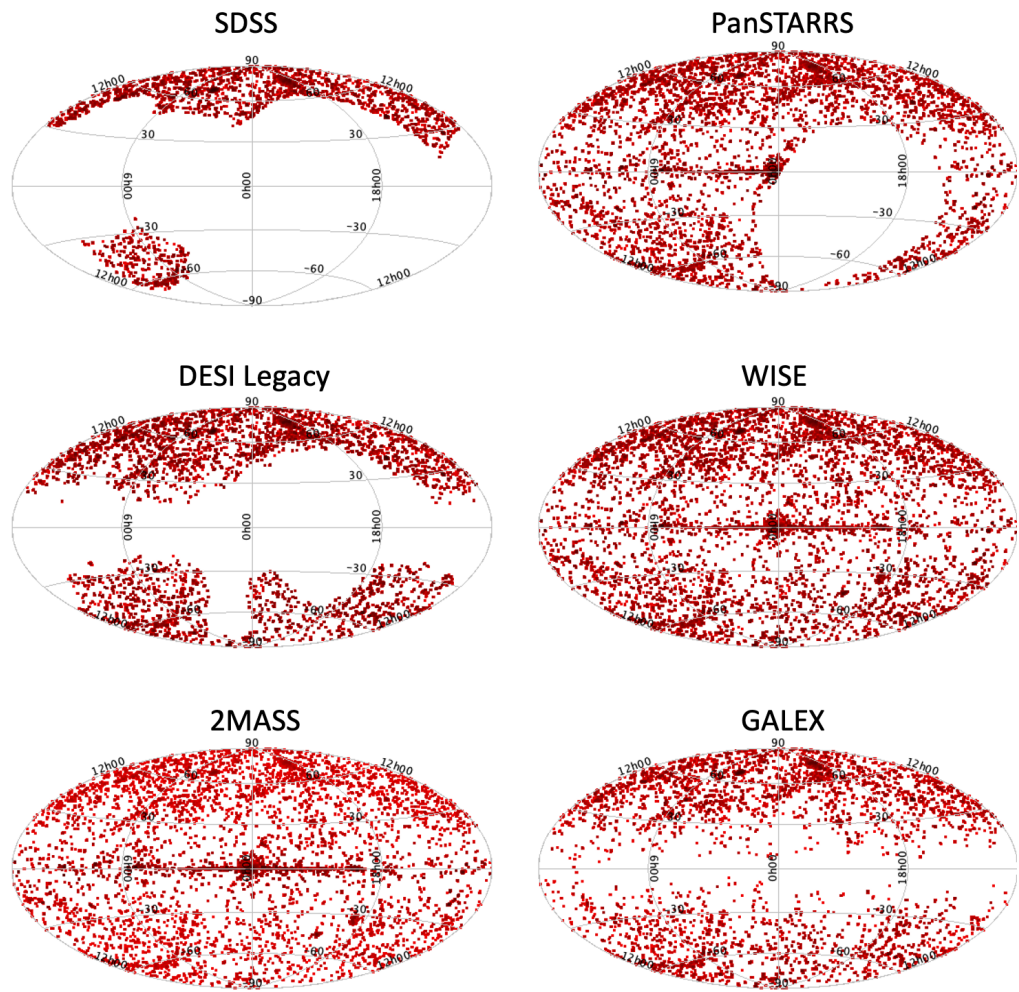

Fig. 2. The optical/IR/UV counterparts of CSC2 sources are plotted in galactic coordinates (l, b).

## 3.4 WISE[8]

The WISE survey covers the entire sky, and the ALLWISE catalog includes 740M sources with near/mid-IR photometric measurements in 4 bands (3.4, 4.6, 12, 22 μm). We used the near/mid-IR colors to classify stars, galaxies, and QSOs,

---

[8] https://wise2.ipac.caltech.edu/docs/release/allsky/

### 3.5 2MASS[9]

The point source catalog includes 470M sources, and the extended source catalog includes 1.6M sources (~70K sources listed only in the extended source catalog). They have photometric measurements in 3 near-IR bands (j, h k).

### 3.6 GALEX[10]

The revised GUV catalog (Bianchi et al. 2017) includes 83M sources with UV photometric data in 2 bands (FUV $\lambda_{eff}$ = 1528A, NUV $\lambda_{eff}$ = 2310A).

## 4. CROSSMATCH

To crossmatch the CSC2 with our suite of reference catalogs, we used the NWAY package v4.5.2[11] (see Salvato et al. 2018 for details). Given that the CSC2 covers a limited part of the sky (~500 deg$^2$), we grouped the stacks[12] into 4380 ensembles, each covering a small part of the continuous sky, each of which we crossmatched individually, with the optical/IR catalog data within the same sky footprint.

The NWAY parameters were set to optimize the matching fraction while minimizing false positives and negatives. (a) We started with a search radius of 10 arcsec, considering the reasonable maximum of CSC2 positional errors to avoid losing any real match, but at the end of the match, we restricted this criterion to accept only unique matches with separation < 3 arcsec, regardless of the positional error. (b) We set the minimum probability ratio for the secondary match *acceptable-prob* = 0.25 - smaller than the default value of 0.5 - to determine a 'unique match' conservatively, particularly in crowded fields with high source densities. (c) We set *match_flag*=1 (indicating the most probable match). (d) We set the probability threshold (i.e., the probability that one of the associations is correct) *p_any* = 0.5. With these rather strict selection criteria, we have minimized the false match rate to ~5%, as described below and in Appendix A.

To evaluate the rate of false matches, we have run extensive simulations, following the matching procedure outlined above but after shifting the source positions in eight directions (horizontal, vertical, and diagonal) by 30 arcseconds. Not surprisingly, we found that the rate of false matches is a strong function of source density. When the source density is low (< 10 sources in arcmin$^2$), the rate of false matches is also low (< 5%). When the source density is intermediate (10 – 30 sources in arcmin$^2$), the rate of false positives is about ~10%. When the source density is very high (> 40 sources in arcmin$^2$), the rate of false positives can be as high as ~40%. We present the complete simulation results in Appendix A.

To minimize the false match rate by avoiding crowded fields, we limited the range of Galactic latitude (|b| > 15) and removed the ensembles with high optical/IR source densities, such as those containing nearby large galaxies (e.g., M31 and M33, LMC, SMC). For ensembles partially containing regions of high source density, such as Galactic globular clusters (e.g., M2, M3, M5) we excised the high-density region. After excluding these crowded fields and applying

---

[9] https://irsa.ipac.caltech.edu/data/2MASS/docs/
[10] http://www.galex.caltech.edu
[11] https://github.com/JohannesBuchner/nway
[12] https://cxc.cfa.harvard.edu/csc/about.html

our strict selection criteria, we achieve a rate of false positive matches of ~5% in our matches between CSC2 and other catalogs.

An assessment of the quality of the crossmatching can be made based on the spatial separation between the optical/IR counterparts for a given X-ray source. Since the positional uncertainties of the optical/IR catalogs are considerably smaller than those of the CSC2 sources (e.g., 0.05″ on average for the SDSS), the separation of the matched counterparts should be small if the match is not due to chance. Among the CSC2 sources with multiple optical counterparts from SDSS, PanSTARRS, and LEGACY catalogs, we find that the separations between the two optical sources exceed 1″ for only 0.4% of the matches. The fraction becomes higher (3%) if 2MASS counterparts are included and is the highest (9%) if WISE counterparts are included. Based on this, we further excluded from the CGC all the CSC2 sources with a large separation between two counterparts (> 1″ without WISE or > 2″ with WISE). This additional condition removes ~1% of the matches for which at least one counterpart is likely false.

## 5. CLASSIFICATION

### 5.1 SDSS Classification and Sample Extraction

Crossmatching CSC2 and SDSS, we found ~34k unique SDSS counterparts (Figure 2). We summarize the number of objects with spectroscopic, photometric, and morphological data in Table 1. From these matches we have defined two high quality subsamples, the *spec-z* and the *photo-z* samples, as described below. A detailed justification of our selection criteria is given in Appendix A.

To assemble the *spec-z* sample we have used SDSS counterparts with spectroscopic information with *zWarning* = 0, to exclude low-quality spectroscopic redshifts. We also excluded a small number of objects with large errors ($error(z) / z > 0.5$; see Bolton et al. 2012 for the list of SDSS quality checks). After removing ~600 matches with low-quality data, we have 9350 objects with high-quality *spec-z* and spectral classes.

For the *photo-z* sample, we use the photometric redshifts (*photo-z*) from Beck et al. (2016). They used multiple training sets from optical galaxy catalogs and applied various galaxy templates in SED fitting. As their sample effectively excluded QSOs and stars, our photo-z sample primarily consists of galaxies (see section 5.1.3). From this sample, we selected only objects with *photoErrorClass* = 1 (Beck et al. 2016). We also exclude a small number of objects with $error(z) / z > 0.5$. This selection results in the removal of ~60% of the *photo-z* SDSS counterparts leaving 7617 objects with high-quality *photo-z*. Of these, 4834 are not included in the SDSS *spec-z* sample.

```
Table 1. SDSS counterparts
───────────────────────────────────────────────────────────────────
Total number of matches                           34085
Objects with high-quality spec-z                  9350*
Objects with high-quality photo-z (without Spec-z)  7617 (4834+)
Objects with morphological type from GalaxyZoo    1249
───────────────────────────────────────────────────────────────────

* SDSS spec-z sample
+ SDSS photo-z sample
```

The spectroscopic classification of the *spec-z* sample identifies 22% of the sample as galaxies, 77% as QSOs, and 0.4% as stars (see Table 2). Since we cannot determine the distances of stellar CSC2 sources in the Milky Way from the SDSS information, we will exclude the CSC2 sources with 'star' counterpart from most of the considerations of this paper, but we will consider this sample for future work.

Of the galaxies, 23% show star formation and ~60% are classified as unknown (NaN), based on the SDSS subclass[13]. The latter likely includes a large number of early-type galaxies with no detectable emission lines. The remaining 18% are assigned as subclass = Broadline or AGN. We checked whether these objects with some AGN signatures differ from the other spectroscopically classified galaxies in their $L_X$, $F_{XO}$, and WISE colors and found no significant difference (See Figure 2 and 13 in Kim et al. 2023). Since we primarily rely on $L_X$, $F_{XO}$, and WISE colors to select galaxies, we consider the entire 2071 objects with class = galaxy as the spectroscopically classified galaxies. Also, we include the sub-class in our catalog so that users can apply their own selections for their specific goals. We note that we will continue to investigate multiwavelength properties (from radio to X-ray) in different types of galaxies (early type, late type, AGN, LINERS, etc.) in forthcoming papers.

Among the spectroscopically classified QSOs, three-quarters exhibit broad emission lines, while 5% are star-forming and 17% are unknown. The spectroscopically classified stars also have subclasses (e.g., stellar spectral types), but we do not list them here.

Table 2. SDSS spectroscopic subclass

| subclass | Galaxy | QSO | Star |
|---|---|---|---|
| total | 2071 | 7240 | 39 |
| Broadline | 93 | 5571 | |
| Starburst | 125 | 316 | |
| Starforming | 349 | 23 | |
| AGN | 293 | 80 | |
| NaN | 1211 | 1255 | |

### 5.1.1 X-ray to Optical Flux Ratio ($F_{XO} = F_X/F_O$) Distributions

We define the X-ray to Optical flux ratio $F_{XO}$, as in Maccacaro et al. (1988): Log ($F_{XO}$) = log ($F_X$) + 5.31 + r / 2.5, and calculated $F_{XO}$ values for the matched sources from the broad-band (0.5-7 keV) X-ray flux from CSC2 and the r-band magnitude in the SDSS *PhotoObj* table. $F_{XO}$ has been used in the literature to discriminate between galaxies and AGNs (e.g., Maccacaro et al. 1988, Kim et al. 1992, 2006; Shapley et al. 2001). Most AGNs have $F_{XO}$ values within the range of 0.1-10, while galaxies dominate at lower $F_{XO}$ (< 0.1).

In Figure 3 (left), we plot the r magnitude vs. the X-ray flux ($F_X$) in the *spec-z* sample. The points are color-coded to show the SDSS spectroscopic class. We overplot lines representing three different $F_{XO}$ values: 0.01, 0.1, and 10. At lower $F_{XO}$ (< 0.01) the distribution is overwhelmingly dominated by galaxies (red points), but objects spectroscopically classified as galaxies are also found at higher values of $F_{XO}$, overlapping the distribution of QSOs (blue points).

---
[12] https://www.sdss.org/dr16/spectro/catalogs/#Objectinformation

In Figure 3 (right), we show the histogram of $F_{XO}$ for galaxies, QSOs, and stars. QSOs fall within the $F_{XO}$ range of 0.1-10, as expected. Galaxies are in the $F_{XO}$ range of $10^{-4}$ - 10. Stars are mixed with galaxies at $F_{XO}$ (0.001 – 1). Typical late-type stars (K, M) have $F_{XO}$ in this range (e.g., Zombeck 1990).

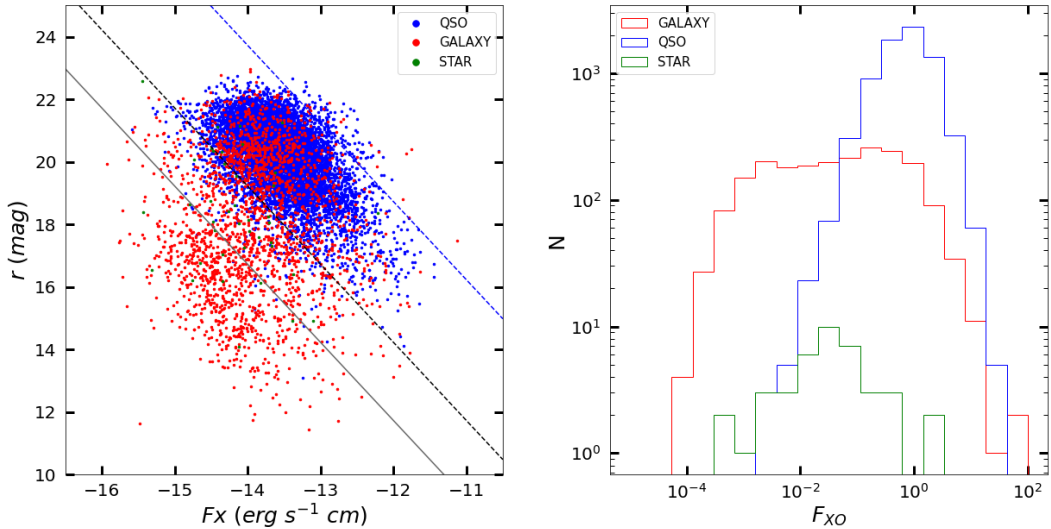

Fig. 3. (Left) Optical r mag against X-ray flux for the SDSS *spec-z* sample. The diagonal lines indicate $F_{XO}$ values of 0.01, 0.1, and 10 from the lower left to the upper right. (Right) $F_{XO}$ distribution for the QSO, galaxy, and star subsamples.

Table 3 lists the fraction of each class in different $F_{XO}$ bins. In the lowest $F_{XO}$ bin ($F_{XO}$ < 0.01), most sources (94%) are spectroscopically classified galaxies; there are 671 galaxies in this bin, ~1/3 of 2071 spectroscopically classified galaxies. In the intermediate $F_{XO}$ bin (0.01 - 0.1), galaxies are the majority (60%), but QSOs account for one-third of the objects. In the highest $F_{XO}$ bin ($F_{XO}$ > 0.1), QSOs dominate (~ 90%).

Table 3. Number of objects in different $F_{XO}$ bins for the SDSS samples

| Fxo | spec-z sample | | | | photo-z sample | |
|---|---|---|---|---|---|---|
| | Galaxies | QSOs | Stars | total | total | galaxies* |
| - 0.01 | 671 (97%) | 10 ( 2%) | 10 ( 1%) | 691 | 252 | 244** |
| 0.01 - 0.1 | 533 (61%) | 323 (37%) | 21 ( 2%) | 877 | 905 | 550 |
| 0.1 - | 867 (11%) | 6907 (89%) | 8 ( 0%) | 7782 | 3677 | 410 |
| Total | 2071 | 7240 | 39 | 9350 | 4834 | 1204 |

* Expected number of galaxies in each $F_{XO}$ bin from the photo-z sample
** Number of galaxies without significant contamination

We estimated the number of galaxies we expect in the *photo-z* sample, assuming the same galaxy fraction found in the *spec-z* sample. These estimates are listed in the last two columns of Table 3. We expect ~1204 galaxies in the photo-z sample. However, there are only 244 galaxies in the lowest $F_{XO}$ bin, where we can identify galaxies with high confidence of 97% (see Table 3).

### 5.1.2 Redshift and X-ray Luminosity ($L_X$) Distributions

In Figure 4, we show the redshift and $L_X$ distributions of each class in the *spec-z* sample. Most galaxies in our sample are within the redshift range z = 0.01-1, while most QSOs are at higher redshifts, z = 0.1-10. QSOs dominate at $L_X > 10^{43}$ erg s$^{-1}$. At $L_X < 10^{43}$ erg s$^{-1}$, galaxies are most abundant.

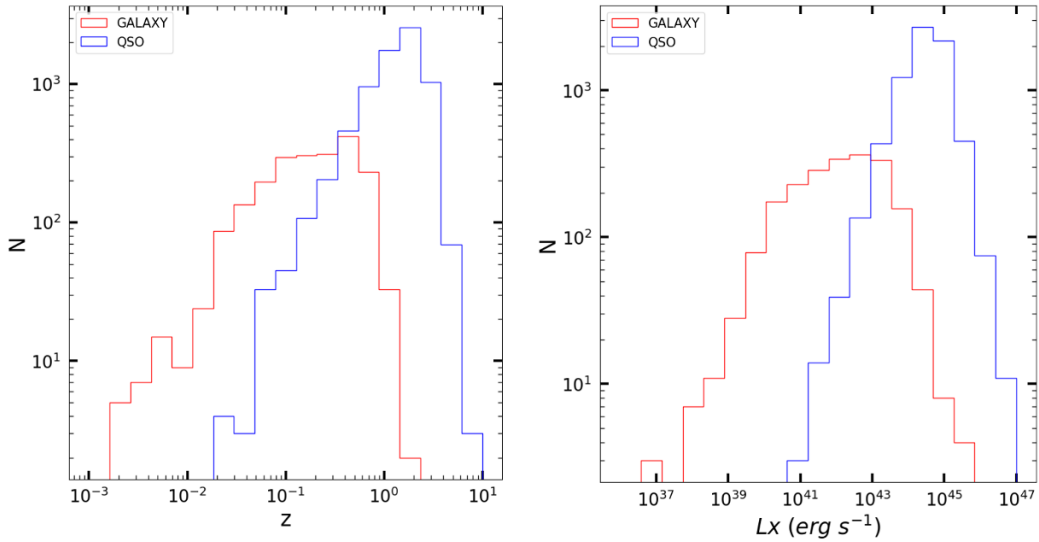

Fig. 4. The redshift and $L_X$ distributions of galaxies and QSOs for the SDSS spec-z sample.

Table 4 lists the fraction of each spectroscopically classified object (galaxy and QSO) in the spec-z sample in two $L_X$ bins. In the low $L_X$ bin ($< 10^{42}$), galaxies are the majority (98%). This is consistent with what was shown by observations of the near universe. For this reason, $L_X = 10^{42}$ erg s$^{-1}$ has been used as a discriminator between galaxies and AGNs (e.g., Moran et al. 1999; Kim et al. 2019). However, only 45% of the spectroscopically classified galaxies (945 out of 2071) are found in this $L_X$ bin, as also shown in Figure 4. In the high $L_X$ bin ($L_X > 10^{42}$), QSOs dominate (~ 87%), but this bin also contains 55% of the CSC2 sources with galaxy spectral classification.

We estimated the number of galaxies we can expect from the *photo-z* sample based on the galaxy fraction in the *spec-z* sample. See the two last columns in Table 4. In total, we expect to have about 1292 galaxies from the photo-z sample. However, there are only 742 galaxies in the low $L_X$ bin, where we can identify galaxies with high confidence of 98% (see Table 4).

For completeness, we describe in Appendix F the redshift and $L_X$ distributions of different samples, including the spec-z sample from SDSS and the photo-z samples from SDSS, PanSTARRS, and Legacy. They do not vary significantly from one sample to another.

Table 4. Number of objects in different $L_X$ bins for the SDSS samples

| $L_X$ (erg/s) | spec-z sample | | | photo-z sample | |
|---|---|---|---|---|---|
| | Galaxies | QSOs | Total | total | galaxies* |
| - 1e42 | 945 (98%) | 24 ( 2%) | 969 | 757 | 742** |
| 1e42 - | 1126 (13%) | 7216 (87%) | 8342 | 4077 | 550 |
| Total | 2071 | 7240 | 9311 | 4834 | 1292 |

\* Expected number of galaxies in each LX bin from the photo-z sample
\*\* Number of galaxies without significant contamination

### 5.1.3 Using $F_{XO}$ and $L_X$ Together as Classifiers

Here we show that utilizing both $F_{XO}$ and $L_X$ simultaneously provides a more efficient way to identify galaxies. This method can be applied to the SDSS *photo-z* sample and to the PanSTARRS and Legacy catalogs crossmatches, where no spectroscopic classification is available (see Sections 7 and 8). Figure 5 (top panel) shows the distribution of the *spec-z* sample in the $F_{XO}$ - $L_X$ plane. As in Figure 3, QSOs and galaxies are color-coded. It is obvious that the two classes are better separated in the $L_X$ - $F_{XO}$ plane than by means of $L_X$ or $F_{XO}$ alone. The spectroscopically classified galaxies (red points in Figure 5) lie in a narrow strip with a slope of ~1 in the $L_X$ - $F_{XO}$ plane. The best-fit relation (the red line in the top panel of Figure 5) gives:

$$(L_X / 2.8 \times 10^{42} \text{ erg s}^{-1}) = (F_{XO} / 0.1)^{1.02} \quad \text{for galaxies in the \textit{spec-z} sample.}$$

Throughout this paper, we use the Bayesian approach to linear regression (Kelly 2007). The linear $L_X$ - $F_{XO}$ relation runs for about five orders of magnitude in both $L_X$ ($10^{39} - 10^{44}$) and $F_{XO}$ ($10^{-4}$ - 10). The width of the strip is narrow, less than 1 dex, regardless of $L_X$. This is primarily because the dynamic range of the optical stellar luminosity ($L_r$) is narrow, with 90% of galaxies having $L_r$ in a range of 20 ($10^{10} - 2 \times 10^{11}$ $L_{r\odot}$, see the bottom panel of Figure 5), to be compared with a range of $10^4$ in $L_X$ ($10^{40} - 10^{44}$ erg s$^{-1}$)[14]. While the upper limit of $L_r$ is likely intrinsic, the lower limit is partly due to the selection effect that small galaxies are undetected in X-ray unless they are nearby. Nonetheless, the extensive dynamical range of $L_X$ means that $L_X$ is practically independent of $L_r$ in our sample.

In contrast to the galaxy sample, QSOs are clustered at the upper right corner in the $F_{XO}$ - $L_X$ plane with the highest $L_X$ and $F_{XO}$. There is no correlation between $L_X$ and $F_{XO}$ among the QSO sample. Instead, QSOs show a correlation between $L_X$ and $L_r$ (bottom right panel of Figure 5). This can be understood because the strong nuclei enhance their fluxes both in the optical and X-ray bands. The best-fit relation (the blue line in the bottom panel of Figure 5) is given by

$$(L_X / 1.6 \times 10^{43}) = (L_r / 10^{10})^{0.72} \quad \text{for QSOs in the } spec\text{-}z \text{ sample}$$

---

[14] We determine the r-band luminosity ($L_r$) in units of solar luminosity, assuming the absolute solar magnitude $M_{r\odot}$= 4.68 mag.

This relation is close to the well-known $L_X$-$L_{UV}$ correlation of (unabsorbed) QSOs with a slope of $0.6 \pm 0.1$ (Tananbaum et al. 1979).

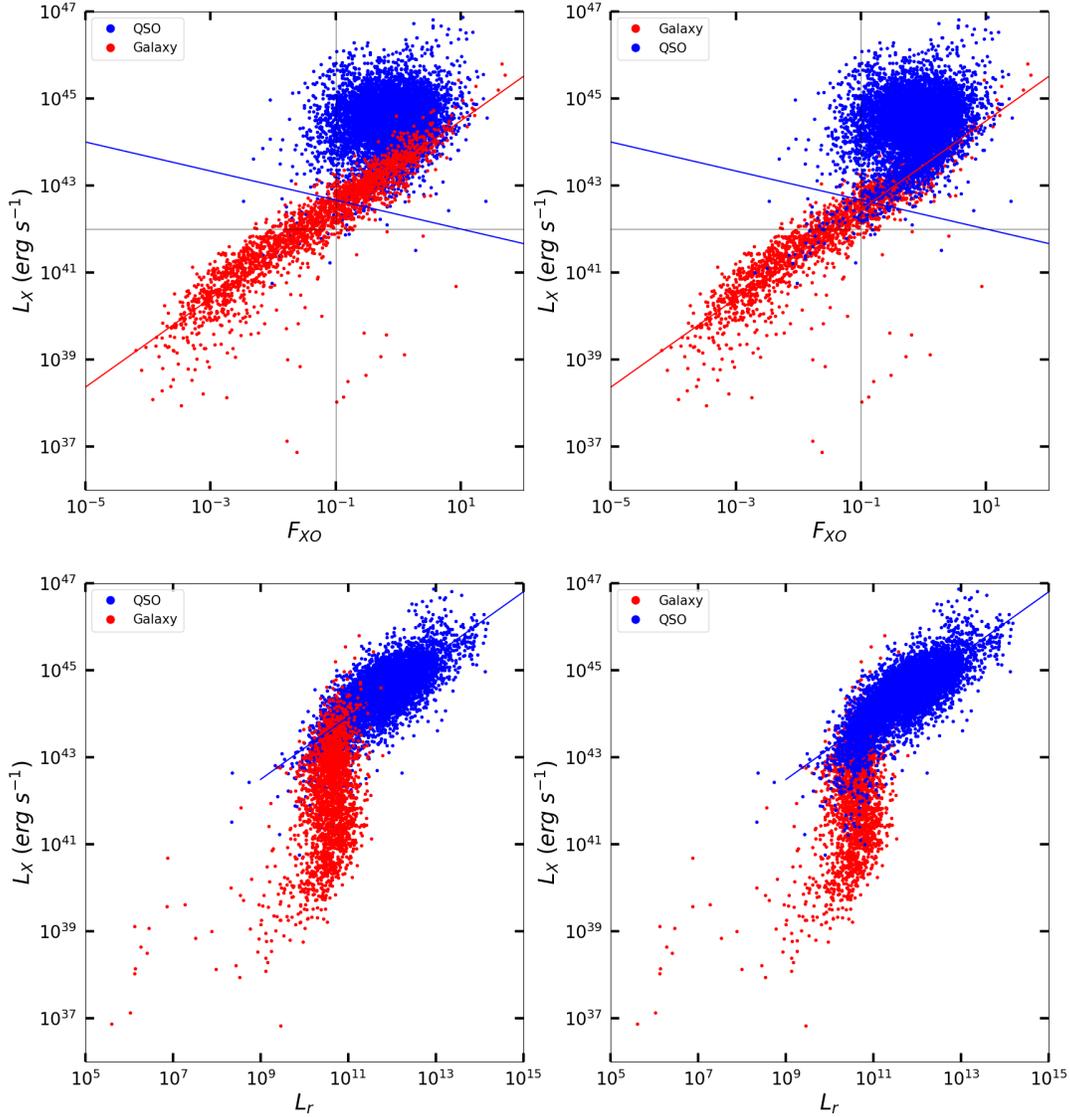

Fig. 5. (top) $L_X$ against $F_{XO}$ for galaxies and QSOs from the spec-z sample. The left and right panels are the same, except (left) galaxies on top of QSOs or (right) QSOs on top of galaxies to visualize both subsamples clearly. The red dashed line is a linear relation that fits the galaxy sample. The blue line is to separate galaxies from QSOs. (bottom) the same but with $L_X$ against $L_r$. The blue line is the best-fit relation for QSOs.

To identify galaxies with the least contaminations by QSOs, we empirically draw a boundary line in the $L_X$ - $F_{XO}$ plane, marked by the blue line in the top panel of Figure 5:

$$(L_X / 4.64 \times 10^{42}) = (F_{XO} / 0.1)^{-1/3} \quad \text{the galaxy – QSO boundary}$$

In Table 5, we list the number of galaxies in the two regions (1) above (2) below the galaxy–QSO boundary line. From region 2, we recover 1311 galaxies (63%) out of 2071 spectroscopically classified galaxies. This fraction is considerably higher than that by $F_{XO}$ alone (1/3 – see section 5.1.1) and that by $L_X$ alone (1/2 – see section 5.1.2). The QSO contamination in this region is 6%.

Table 5. Number of objects in different $F_{XO}$ and $L_X$ bins for the SDSS samples

| ($F_{xo}$, $L_x$) | spec-z sample | | | photo-z sample | |
|---|---|---|---|---|---|
| | Galaxies | QSOs | Total | total | galaxies* |
| region 1 | 760 (10%) | 7151 (90%) | 7911 | 3192 | 307 |
| region 2 | 1311 (94%) | 89 ( 6%) | 1400 | 1642 | 1540 ** |
| Total | 2071 (22%) | 7240 (78%) | 9311 | 4834 | 1847 |

(region 1) above the galaxy–QSO boundary line
(region 2) below the galaxy–QSO boundary line
* Expected number of galaxies in each ($F_{XO}$, $L_X$) bin from the photo-z sample
** Number of galaxies without significant contamination

The missing galaxies (37% of the total spectroscopically classified galaxies) are in the high $L_X$ and $F_{XO}$ region. These X-ray bright galaxies have been called in the literature 'X-ray bright optically normal galaxies' (XBONG, e.g., Elvis 1981; Fiore et al. 2000). We will explore this unusual type of object in a separate paper (Kim et al. 2023 in prep.)

Figure 6 compares the distributions of *photo-z* and *spec-z* samples in the $L_X$ – $F_{XO}$ plane (the top panel) and the $L_X$ – $L_r$ plane (the bottom panel). The distribution of the *photo-z* sample (the red points in Figure 6; as discussed in Section 5.1 this sample primarily consists of galaxies) has a virtually identical slope to that of the spectroscopically classified galaxies (the red points in Figure 5) but a 0.1 dex lower intercept. The best-fit relation of the photo-z sample is

$(L_X / 2.2 \times 10^{42}) = (F_{XO} / 0.1)^{1.01}$   for the *photo-z* sample

and extends to the XBONG regime as observed in the *spec-z* sample.

As in sections 5.1.1 and 5.1.2, we also estimate the number of galaxies expected from the photo-z sample. See the last two columns in Table 5. From the *photo-z* sample, the total number of objects in region 2 between the two boundary lines is 1642. We expect 1540 galaxies with a contamination fraction of 6%. However, the contamination is expected to be lower, given that the photo-z sample preferentially includes galaxies.

In summary, we find 2071 galaxies from the SDSS *spec-z* sample and identify 1642 galaxies candidates from the SDSS *photo-z* sample. In total, we obtain an X-ray galaxy catalog with 3713 galaxy candidates with 102 (= 1642 – 1540) non-galaxies, i.e., a contamination fraction of 3%.

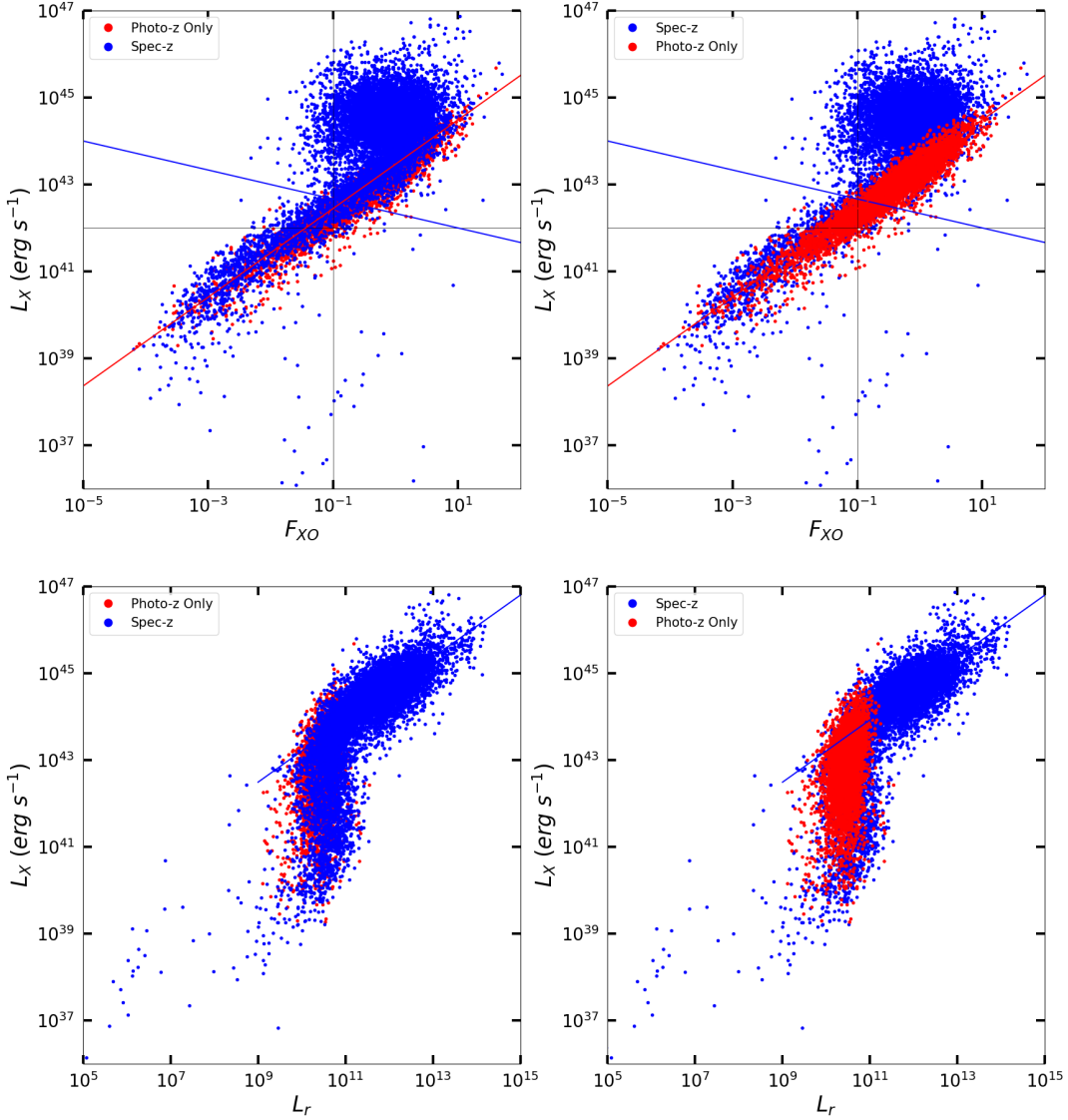

Fig. 6. $L_X$ against (top) $F_{XO}$ and (bottom) $L_r$ from spec-z and photo-z samples. The left and right panels are the same, except (left) the spec-z sample on top of the photo-z sample or (right) vis versa to visualize both samples clearly. The diagonal lines are the same as in Figure 5.

## 5.2 CSC2-SDSS-WISE Sample Selection

In addition to $F_{XO}$ and $L_X$, we have used the WISE colors to identify galaxies and exclude QSOs. Different types of objects occupy unique positions in the IR color-color space (e.g., Wright et al. 2010, Jarrett et al. 2017). We have cross-matched the CSC2 and WISE (All-WISE) catalogs with the same method described in Section 4, finding 124,889 matches. Because the optical information is necessary for the optical magnitude and redshift, from these matches we selected 85,314 CSC-WISE matches with optical counterparts in the SDSS, PanSTARRS, and Legacy catalogs. We then applied a set of quality control criteria and identify 25,190 WISE counterparts. These criteria include limiting galactic latitude (|b|>15 as in Sections 2.1 and 4), excluding well-known crowded

fields (Sections 2.1 and 4), limiting the separations between optical/IR counterparts (<2 arcsec; Section 4), and removing objects with low-quality z data (Section 5.1). Furthermore, we selected sources with S/N > 2 in two and three of the WISE bands, W1 (3.4 µm), W2 (4.6 µm), and W3 (12 µm), as listed in Table 6. We have 13,627 WISE-CSC matches with S/N > 2 in three bands. Of these, 7956 have redshifts from SDSS, 5482 with *spec-z*, and 2474 with *photo-z*. We call them the WISE *spec-z* and *photo-z* samples, respectively.

Table 6. CSC-SDSS-WISE matches

| CSC-WISE | all | selected* | SDSS spec-z | SDSS photo-z (without spec-z) |
|---|---|---|---|---|
| WISE detected | 85314 | 25190 | 7991 | 4396 |
| S/N (W1,W2)>2 | 82181 | 24684 | 7920 | 4355 |
| S/N (W1,W2,W3)>2 | 41177 | 13627 | 5482 ^ | 2474 + |

\* Selected e.g., by galactic latitude, z quality, and crowd fields (see text).
^ WISE spec-z sample
+ WISE photo-z sample

### 5.2.1 WISE IR Colors Classification

Figure 7 plots the WISE *spec-z* and *photo-z* samples in the $W_{12}$ - $W_{23}$ color-color diagram, where $W_{12}$ = W1-W2 and $W_{23}$ = W2-W3. The IR magnitudes are in the Vega magnitude system. The top panel of Figure 7 shows the WISE *spec-z* sample (5480 objects) - spectroscopically classified galaxies and QSOs color-coded, as in Figure 5. QSOs are clustered at higher $W_{12}$ (0.5-1.7) and $W_{23}$ (2-4) colors. In contrast, galaxies are at lower $W_{12}$ colors (0-1) but at a wide range of $W_{23}$ (0-4.5). The previous AGN surveys often applied a lower limit of $W_{12}$ to identify QSOs. For example, Jarett et al. (2011) and Stern et al. (2012) applied $W_{12} > 0.8$, while Assef et al. (2013) applied a similar but W2-dependent limit. The blue dashed line in Figure 7 indicates $W_{12} = 0.8$. As seen in the figure, this limit is helpful in selecting QSOs with a small galaxy contamination (3%). Under this limit ($W_{12} < 0.8$), galaxies dominate, but the QSO fraction is still high. To effectively select galaxies by reducing the QSO contamination, we apply a strict upper limit of $W_{12} < 0.4$ for galaxies. The number of objects in different $W_{12}$ bins is summarized in Table 7. In the lowest $W_{12}$ bin (< 0.4), 93% of objects are galaxies with 7% contamination by QSOs. In the middle bin ($W_{12}$ = 0.4-0.8), galaxies and QSOs are mixed roughly by a ratio of 1:2.

In the bottom panel of Figure 7, the WISE *spec-z* and *photo-z* samples are compared, color-coded as in the bottom panel of Figure 6. In contrast to the $F_{XO} – L_X$ plot in Figure 6, where the distribution of the *photo-z* sample resembles that of the *spec-z* galaxy sample, the WISE *photo-z* sample falls in between the *spec-z* galaxy sample and QSO samples, indicating that galaxies and QSOs are mixed in the WISE *photo-z* sample. Assuming the same galaxy fraction in each $W_{12}$ bin, we expect that 46% (1142 out of 2474) of the WISE photo-z sample will be galaxies (see the last two columns in Table 7). In the first bin ($W_{12} < 0.4$), where galaxies dominate, we can identify 760 galaxy candidates with 7% QSO contamination.

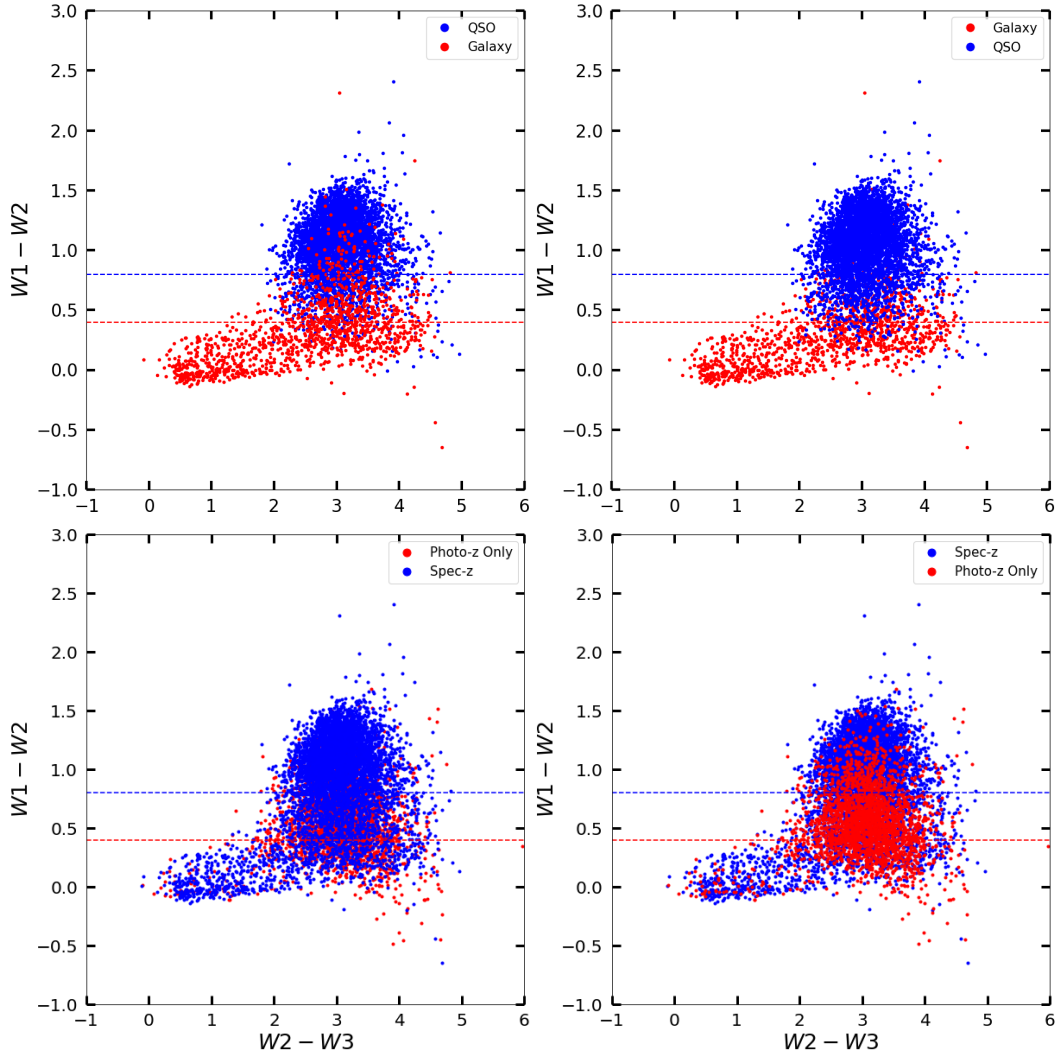

Fig. 7. Top: WISE *spec-z* color-color plot for galaxies and QSOs. Bottom: comparison of WISE *spec-z* and *photo-z* samples. The left and right panels are the same, except the different objects are drawn on top to visualize both samples clearly. The blue dashed line is for $W_{12} = 0.8$, often used to identify QSOs without galaxy contaminations. The red dashed line is for $W_{12} = 0.4$, which we apply to select galaxies without QSO contaminations.

Table 7. Number of objects in different $W_{12}$ (W1-W2) bins for the SDSS samples

| $W_{12}$ (W1-W2) | spec-z sample | | | photo-z sample | |
|---|---|---|---|---|---|
| | Galaxies | QSOs | Total | total | galaxies* |
| - 0.4 | 858 (93%) | 68 ( 7%) | 926 | 760 | 704 ** |
| 0.4 - 0.8 | 382 (35%) | 720 (65%) | 1102 | 1245 | 431 |
| 0.8 - | 90 ( 3%) | 3352 (97%) | 3442 | 469 | 12 |
| total | 1330 (24%) | 4140 (76%) | 5470 | 2474 | 1142 |

* Expected number of galaxies in each W12 bin from the photo-z sample
** Number of galaxies without significant contamination

### 5.2.2 $L_X$ - $W_{12}$ Classification

While $W_{23}$ is a good indicator to separate passive galaxies with low $W_{23}$ colors and star-forming galaxies with high $W_{23}$ colors (e.g., Wright et al. 2010; see also section 6), there is considerable overlap of galaxies and QSOs in $W_{23}$ color (Figure 7). Therefore, we used $W_{12}$ together with $L_X$ to explore a more efficient way to separate galaxies, QSOs, and sources with stellar SDSS counterparts (Figure 8). We selected a boundary line in the $L_X$ – $W_{12}$ plane to separate galaxies from QSOs, shown in Figure 8 by a blue line:

```
log (Lx) = -2.22  W12 + 43.78      the galaxy – QSO boundary
```

In Table 8, we list the number of galaxies in the two regions (1) above (2) below the galaxy–QSO boundary line. From region 2, we recover 954 galaxies (72%) out of 1330 spectroscopically classified galaxies. The non-galaxy contaminations in these regions are only 4%, all being QSOs. When compared to the efficiency of finding galaxies in the $L_X$-$F_{XO}$ plane (Figure 6), the $L_X$-$W_{12}$ plane provides a slightly more effective result (72% vs. 64%). The contamination fraction is also slightly improved (4% vs. 6%).

The missing galaxies (28% of the spectroscopically classified galaxies) are found in region 1, with the highest $L_X$ and $W_{12}$. These X-ray bright galaxies correspond to XBONGs; we will explore them in a separate paper (Kim et al. 2023 in prep.)

In the bottom panel of Figure 8, the WISE *photo-z* sample (2474 objects) is compared with the WISE spec-z sample (5482 objects). The last two columns in Table 8 show the number of galaxies expected in the *photo-z* sample, based on the spectroscopically classified galaxies. In the *photo-z* sample, the total number of objects in region 2 between the two boundary lines is 944. We expect 902 galaxies with a contamination fraction of 4%, as estimated from the WISE *spec-z* sample.

In summary, we find 2071 Chandra-detected galaxies in the SDSS *spec-z* sample. Additionally, in the SDSS *photo-z* sample, we identify 1642 and 944 galaxies candidates using the $L_X$-$F_{XO}$ selection and the $L_X$-$W_{12}$ selection, respectively; 734 galaxies are found by both selection methods. In total, we obtain an X-ray galaxy catalog with 3939 galaxy candidates with a contamination fraction of 3%.

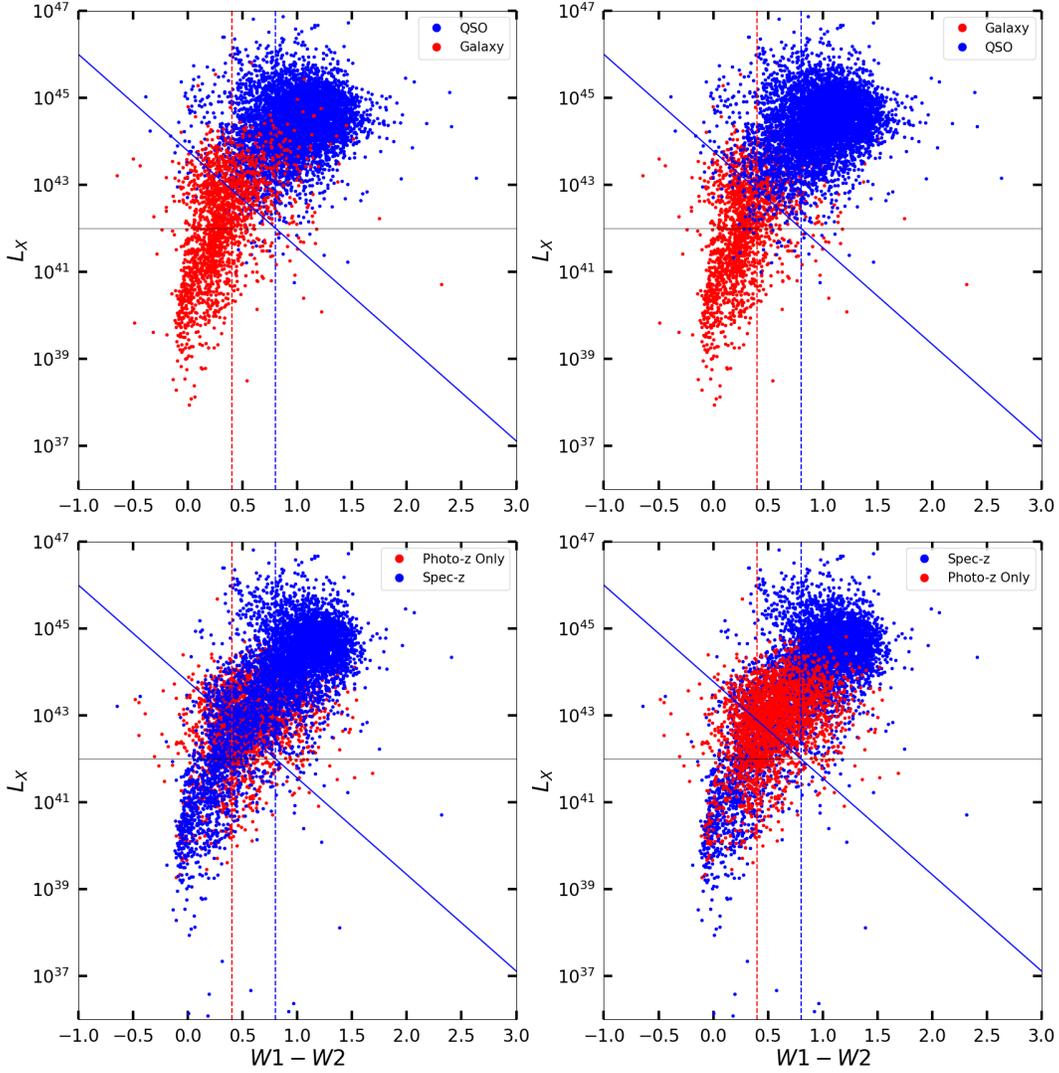

Fig. 8. $L_X$ against WISE color $W_{12}$ (= W1-W2) (top) for galaxies and QSOs from the SDSS spec-z sample (bottom) from the SDSS spec-z and photo-z samples. The left and right panels are the same, except the different objects are drawn on top to visualize both samples clearly. The vertical lines are the same as in Figure 7. The blue line separates galaxies and QSO.

Table 8. Number of objects in different $W_{12}$ and $L_X$ bins for the SDSS samples

| W1-W2 | spec-z sample | | | photo-z sample | |
| --- | --- | --- | --- | --- | --- |
| | Galaxies | QSOs | Total | total | galaxies* |
| region 1 | 376 ( 8%) | 4096 (92%) | 4472 | 1530 | 129 |
| region 2 | 954 (96%) | 44 ( 4%) | 998 | 944 | 902 ** |
| total | 1330 (24%) | 4140 (76%) | 5470 | 2474 | 1031 |

```
__________|________________________________|____________________
```
(region 1) above the galaxy–QSO boundary line
(region 2) below the galaxy–QSO boundary line

 \* Expected number of galaxies in each W12 bin from the photo-z sample
\*\* Number of galaxy candidates without significant contamination

### 5.3 Pan-STARRS and Legacy Crossmatch Samples

Given that the sky coverage of SDSS is limited, we also used two additional wide-area optical survey catalogs:

(1) Pan-STARRS (Flewelling et al. 2020), which covers three-quarters of the entire sky ($\delta > -30°$). See Figure 2 for the sky maps. In the following, we use the abbreviation PS to refer to the Pan-STARRS DR2 catalog. Since PS has no spectroscopic information, we use the photo-z from Beck et al. (2021), who applied a machine learning neural network algorithm trained with a spectroscopic sample of redshifts and classification information to determine a probabilistic photometric classification: Galaxy, Star, QSO, or Unsure, and photometric redshifts for sources identified as galaxies. In Appendix B, we show that these photo-z estimates are consistent with the SDSS spec-z and photo-z. For the r magnitude, we take the Kron mag because it is suitable to measure the total flux of an extended source, such as a galaxy (Kron 1980). If not available, we take the AP mag (Flewelling et al 2020). We validate these choices by comparing r magnitudes for the objects common in SDSS and PS (Appendix C).

(2) The DESI Legacy catalog (Dey et al. 2019). The Legacy South covers the sky area outside the SDSS footprint, while the Legacy North footprint significantly overlaps with that of SDSS (see Figure 1). In this work, we only utilize the Legacy South data. We use the abbreviation LS to refer to the Legacy South DR8 catalog. We take the photo-z from Zou et al. (2019). In Appendix B and C, we show that the LS photo-z and r mag are consistent with those of SDSS and PS.

Applying the crossmatch procedure described in Section 4, we find 105,728 (69,478) PS (LS) counterparts to the CSC2 sources (Table 9), which are reduced to 22,617 (20,204) PS (LS) objects with the set of quality selection criteria described in Section 5.2.

Further excluding those already matched with SDSS objects, we find 8762 PS counterparts: the 'PS sample'. Of these, 3396 are also detected with S/N > 2 in the three WISE bands W1, W2, and W3, which we call it the 'PS-WISE sample'.

Similarly, further excluding those already matched with SDSS and PS objects, we find 7642 LS counterparts, constituting the 'LS sample'. Of these, 2259 are detected with S/N > 2 in the WISE three bands W1, W2, and W3: the 'LS-WISE' sample.

All the objects in the PS and LS samples are photometrically classified as galaxies by Beck et al. (2021) and Zou et al. (2019), who apply galaxy templates to derive photometric redshifts. Figure 9 shows the PS and LS samples in the $L_X$-$F_{XO}$ plane. Note that their distributions are similar to those of the SDSS galaxies (the red points in Figure 5). The red diagonal line is the best fit (with slope 1.02) from the SDSS *spec-z* galaxy sample. This line well represents both PS and LS samples. The best slopes for the PS and LS samples are 0.98 and 0.96, respectively.

Table 9. Pan-STARRS and Legacy counterparts

|  | Pan-STARRS | Legacy |
|---|---|---|
| Total number of matches | 105728 | 69478 |
| selected ** | 22617 | 20204 |
| used here | 8762 * | 7642 + |
| with WISE detection | 3396 ^ | 2259 # |

** selections by galactic latitude, z quality, and crowd fields (see text).
* PS sample (do not have SDSS counterparts)
+ LS sample (do not have SDSS and Pan-STARRS counterparts)
^ PS-WISE sample (snr > 2 in W1, W2, and W3)
# LS-WISE sample (snr > 2 in W1, W2, and W3)

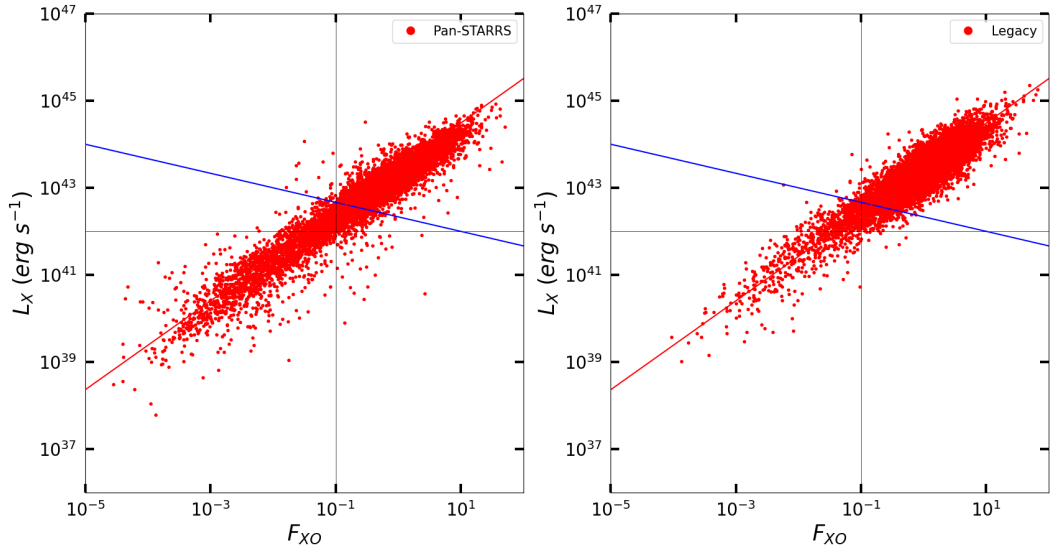

Fig. 9. $L_X$ against $F_{XO}$ for (left) the PS and (right) the LS samples. All lines are the same as in the top panel of Figure 6.

The distributions of the PS-WISE and LS-WISE samples in the $W_{12} - W_{23}$ and $L_X$-$W_{12}$ planes (Figures 10 and 11) are similar to those of the SDSS *photo-z* sample. In Tables 10 and 11, we list the number of objects in three regions of the ($L_X$-$F_{XO}$) and ($L_X$-W) planes. Applying the $L_X$ - $F_{XO}$ selection, we identify 3060 and 1121 (total 4181) galaxy candidates from the PS and LS samples, respectively. Applying the $L_X$ - $W_{12}$ selection, we find 1787 and 653 (total 2440) galaxy candidates from the PS and LS samples, respectively. Of these objects, 1455 and 438 (total 1893) satisfy both selection criteria. In total, we find 4708 galaxy candidates with a contamination fraction of ~ 5%.

Table 10. Number of objects in different $F_{XO}$ and $L_X$ bins for the Pan-STARRS and Legacy samples

| (Fxo, Lx) | Pan-STARRS | Legacy |
|---|---|---|
| region 1 | 5702 | 6519 |
| region 2 | 3060* | 1121* |
| Total | 8762 | 7640 |

(region 1) above the galaxy–QSO boundary line
(region 2) below the galaxy–QSO boundary line

* Number of galaxy candidates without significant contamination

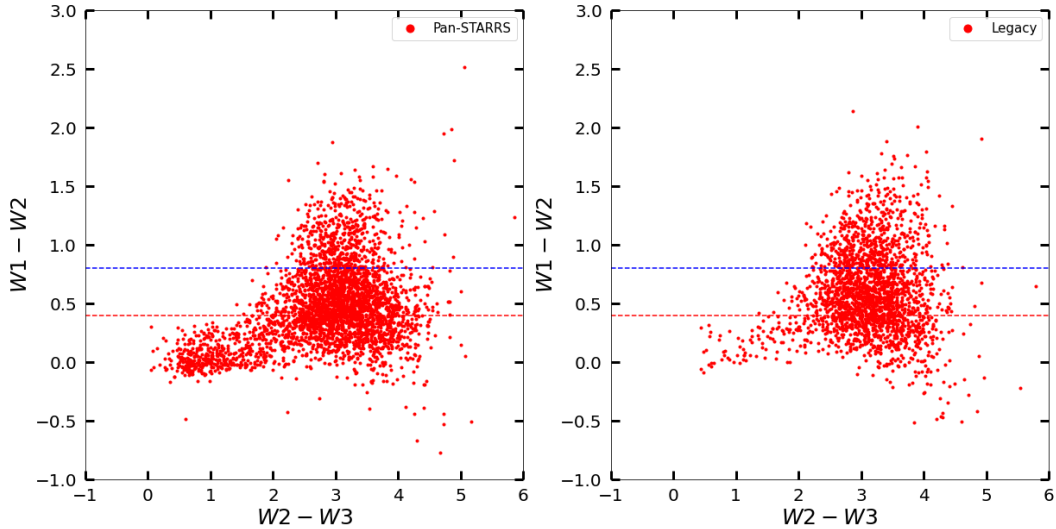

Fig. 10. $W_{12}$ (W1-W2) against $W_{23}$ (W2-W3) for (left) the PS-WISE sample and (right) the LS-WISE sample. All lines are the same as in Figure 7.

In summary, we find 2071 Chandra-detected galaxies from the SDSS *spec-z* sample. Additionally, from the SDSS/PS/LS *photo-z* sample, we identify 5823 and 3384 galaxies candidates using the $L_X$-$F_{XO}$ and the $L_X$-$W_{12}$ selection, respectively; 2611 galaxies are selected by both methods. In total, we select 8667 galaxy candidates with a contamination fraction of ~3%.

We identified 110 nearby galaxies with unidentified off-center X-ray sources in addition to the central source. We exclude them from the final catalog. See Appendix E for detailed descriptions of local galaxies. Additionally, we exclude 10 nearby ($z < 0.01$), optically faint ($Lr < 10^7$ $L_{r\odot}$) objects which are likely globular clusters.

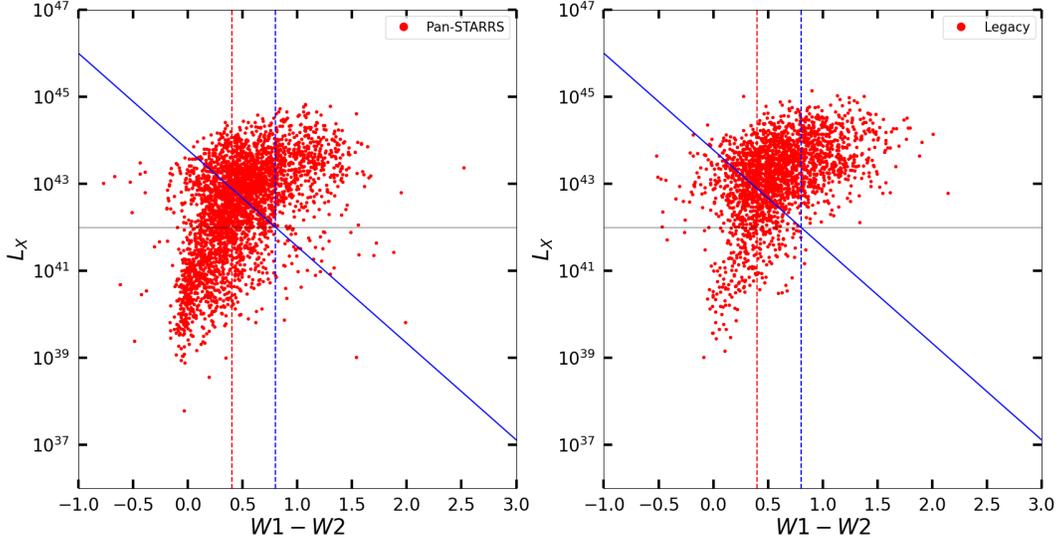

Fig. 11. $L_X$ against $W_{12}$ for (left) the PS-WISE sample and (right) the LS-WISE sample. All lines are the same as in Figure 8.

Table 11. Number of objects in different $W_{12}$ and $L_X$ bins for the Pan-STARRS and Legacy samples

```
_______________________________________________
              |
  (W12, Lx)   |  Pan-STARRS    Legacy
______________|________________________________
              |
   region 1   |      1609        1606
   region 2   |      1787*        653*
              |
______________|________________________________
              |
    total     |      3396        2259
______________|________________________________
```

(region 1) above the galaxy–QSO boundary line
(region 2) below the galaxy–QSO boundary line

\* Number of galaxy candidates without significant contamination

## 6. The CSC2 X-ray Galaxy Catalog

Our final CSC2 Galaxy Catalog (CGC) comprises 8547 galaxies: 24% are selected by optical spectroscopy, 30% by two methods using three photometric quantities ($L_X$, $F_{XO}$, $W_{12}$), and 46% by one method using either ($L_X$-$F_{XO}$) or ($L_X$-$W_{12}$). The fractions of contaminations (primarily by QSOs) are ~3% and ~5% for the selection by ($L_X$-$F_{XO}$) and ($L_X$-$W_{12}$), respectively. The redshift of the CGC galaxies ranges from 0.04 and 0.7 (counting at the 5$^{th}$ and 95$^{th}$ percentiles) with a mean value of 0.3 ± 0.2. The optical luminosity varies from $10^{10}$ to $10^{11}$ $L_{r\odot}$, and the X-ray luminosity ranges from 2 x $10^{40}$ to 2 x $10^{43}$ erg s$^{-1}$ (counting at the 5$^{th}$ and 95$^{th}$ percentiles).

For completeness and convenience, we add the 2MASS near-IR data and GALEX UV data to CGC. The 2MASS and GALEX catalogs are described in section 3, and the same crossmatch procedure in section 4 is applied to find the counterparts.

One galaxy (2CXO J143456.7+340627), classified as a G2 star in the SDSS spectroscopic class, was first rejected in the spec-z sample but re-selected by both $L_X$-$F_{XO}$ and $L_X$-$W_{12}$ methods. We checked the optical images and confirmed that it is a spiral galaxy.

The columns in the CGC are described below, and the online catalog is available on the dedicated website[15] and in a .tar.gz package.

- **CSCID** CSC v2 name like 2CXO_J001625.8-270704. Note that a blank space between 2CXO and J was replaced by '_'
- **SDSSID** SDSS objID
- **PSID** PanSTARRS objID
- **LEGSID** Legacy South ls_id
- **WISEID** WISE designation
- **2MASSID** 2MASS psc/xsc designation
- **GALEXID** GALEX objID
- **ra, dec** J2000 coordinates of the CSC 2 sources
- **sep_X{S, P, L, W}** separations in arcsec between the positions of X-ray and {SDSS, PanSTARRS, Legacy, WISE} sources
- **spec** 1 if the galaxy is selected by the SDSS spectroscopic class, 0 otherwise.
- **photo** 1 if the galaxy with SDSS photo-z is selected by $L_X$-$F_{XO}$, 0 otherwise (see section 5.1.3).
- **wph** 1 if the galaxy with SDSS photo-z is selected by $L_X$-$W_{12}$, 0 otherwise (see section 5.2.2).
- **PS** 1 if the galaxy with PanSTARRS photo-z is selected by $L_X$-$F_{XO}$, 0 otherwise (see section 5.3).
- **wPS** 1 if the galaxy with PanSTARRS photo-z is selected by $L_X$-$W_{12}$, 0 otherwise (see section 5.3).
- **LS** 1 if the galaxy with Legacy South photo-z is selected by $L_X$-$F_{XO}$, 0 otherwise (see section 5.3).
- **wLS** 1 if the galaxy with Legacy South photo-z is selected by $L_X$-$W_{12}$, 0 otherwise (see section 5.3).
- **z** redshift
- **z_error** redshift error
- **z_code** source of the redshift. z_SDSSsp for SDSS spec-z; z_SDSSph for SDSS photo-z; z_PSph for PanSTARRS photo-z; z_LEGSph for Legacy photo-z; z_LEGSsp for Legacy spec-z; 2MRS for 2MASS redshift survey
- **d** luminosity distance in Mpc, assuming $H_o$=69.6 and $\Omega_m$=0.286
- **r** optical magnitude in the r band
- **r_code** source of the r mag. r_SDSS for SDSS; rPSkron for PanSTARRS kron mag; ; rPSap for PanSTARRS aperture mag; r_LEGS for Legacy South
- **g** optical magnitude in the g band
- **g_code** source of the g mag. Same as in r_code.
- **k** 2MASS k-band magnitude

---

[15] https://cxc.cfa.harvard.edu/GalaxyAtlas/CGC/

- **W1** WISE magnitude in the W1-band (3.4 μm)
- **w1snr** W1 signal-to-noise ratio
- **W2** WISE magnitude in the W2-band (4.6 μm)
- **w2snr** W2 signal-to-noise ratio
- **W3** WISE magnitude in the W3-band (12 μm)
- **w3snr** W3 signal-to-noise ratio
- **W12** W1-W2 color
- **W23** W2-W3 color
- **nuv_mag** GALEX magnitude in the near-UV band ($\lambda_{eff}$ = 2310 Å)
- **fuv_mag** GALEX magnitude in the far-UV band ($\lambda_{eff}$ = 1528 Å)
- **significance** source flux significance from the CSC2 master source table
- **likelihood_class** True or Marginal from the CSC2 master source table
- **FxB15** X-ray flux in B-band (0.5-7 keV) in unit of $10^{-15}$ erg s$^{-1}$ cm$^{-2}$
- **FxS15** X-ray flux in S-band (0.5-2 keV) in unit of $10^{-15}$ erg s$^{-1}$ cm$^{-2}$. This is the sum of CSC soft (0.5-1.2 keV) and medium (1.2-2 keV) band fluxes.
- **FxH15** X-ray flux in H-band (2-7 keV) in unit of $10^{-15}$ erg s$^{-1}$ cm$^{-2}$
- **FxB15_lolim** lower limit of FxB15
- **FxB15_hilim** upper limit of FxB15
- **FxB_code** source of the B-band flux from CSC2. flux_aper_b = aperture flux in B-band (95%). flux_aper_a = sum of the S and H fluxes if flux_aper_b is unavailable (5%). flux_aper_w = aperture flux in W band for HRC only (0.1%)
- **FxS_code** source of the S-band flux from CSC2
- **FxH_code** source of the H-band flux from CSC2
- **LxB40** X-ray luminosity in B-band (0.5-7keV) in unit of $10^{40}$ erg s$^{-1}$
- **LxS40** X-ray luminosity in S-band (0.5-2keV) in unit of $10^{40}$ erg s$^{-1}$
- **LxH40** X-ray luminosity in H-band (2-7keV) in unit of $10^{40}$ erg s$^{-1}$
- **Lr10** Optical luminosity in r band in the Solar unit, assuming Mr(sun)= 4.68 mag
- **Lk10** NIR luminosity in k band in the Solar unit, assuming Mk(sun)= 3.28 mag
- **logFxo** log(Fx/Fo) (see section 5.1.1)
- **class_SDSS** SDSS spectroscopic class
- **subclass_SDSS** SDSS spectroscopic subclass
- **class_PS** PanSTARRS class (Galaxy, QSO, Star)
- **type_PS** PanSTARRS type (E for early-type galaxies, S for spiral galaxies)
- **type_LEG** Legacy type (Dev for the de Vaucouleurs profile; EXP for the exponential profile with a variable axis ratio; REX for the round exponential)
- **p_cs** fraction of Galaxy Zoo votes for combined spiral (clockwise + anti-clockwise + edge-on)
- **p_el** fraction of Galaxy Zoo votes for elliptical
- **p_mg** fraction of Galaxy Zoo votes for merger
- **p_dk** fraction of Galaxy Zoo votes for don't know

# 7. Discussions and Future Plans

The CGC provides a basis for the investigation of a number of critical scientific questions. We describe some examples below, including results from previous studies and investigations we plan to publish in the near future.

## 7.1 X-ray scaling relations in early and late-type galaxies

The X-ray emission of early and late-type galaxies has different properties: (1) The X-ray emission from a gravitationally bound hot halo is the dominant source in typical giant elliptical galaxies, which is correlated with $M_{TOTAL}$ (including dark matter) but not with $M_\star$ (or $L_K$). Additionally, low-mass X-ray binaries (LMXBs) whose total $L_X$ is proportional to $M_\star$ or $L_K$ are present in all early-type galaxies and become the dominant source in X-ray faint early-type galaxies (e.g., Boroson, Kim & Fabbiano 2011). (2) The hot ISM and high-mass X-ray binaries (HMXBs), directly related to star formation episodes, dominate in spiral galaxies (e.g., Lehmer et al. 2010; Li & Wang 2013). (3) AGNs (mostly low-luminosity AGNs - LLAGNs) may be present in some galaxies of both types.

In the CGC, we provide available information to separate the CGC galaxies into early and late types. From the SDSS spectroscopy database, we include the subclass - STARBURST, STARFROMING, or unclassified. The unclassified galaxies are likely to be early-type galaxies (ETG) with no strong emission lines. We also include in the CGC the morphological types based on the citizen votes from the Galaxy Zoo catalog and those from the PanSTARRS catalog of broad morphology by Goddard and Shamir (2020. We include from the DESI Legacy catalog, the radial profile types - DEV (de Vaucouleurs profile), EXP (exponential profile with a variable axis ratio), or REX (round exponential). In future work, we plan to use these morphology types and investigate the scaling relations and their implications separately in early and late-type galaxies.

## 7.2 X-ray luminosity function

The X-ray luminosity function (XLF) is a critical measurement that can be used to test theoretical models and numerical simulations. However, XLFs can only be built from large, unbiased samples. Most XLFs published so far are based on a small number of (preferentially X-ray bright) galaxies. For example, Ptak et al. (2007) used 40 early-type and 46 late-type galaxies from the GOODS fields, and Tzanavaris and Georgantopoulos (2008) used 101 early-type and 106 late-type galaxies from the combined data of the Chandra deep fields and XBootes fields. These XLFs are highly unconstrained both at the low and high luminosity end and cannot constrain dependencies on mass or redshift convincingly.

Using the CGC, we can apply the modified 1/VMAX (non-parametric) method and the (parametric) maximum likelihood techniques, with special care of the Eddington bias and the uncertainty in photo-z, e.g., by using deconvolution- or convolution-based estimators. We can compare the observed XLF with cosmological simulations, which so far has only been done for the XRB XLFs (e.g., Fragos et al. 2013).

## 7.3 Rare objects - XBONG

The existence of an unusual population of X-ray bright, optically normal galaxies (XBONGs) has been known since the Einstein mission (Elvis et al. 1981). Their X-ray emission is as luminous as that of typical AGNs ($L_X > 10^{42}$ erg s$^{-1}$), but they show no signs of AGN activity in their optical spectra. XBONGs are attracting more attention on both theoretical and observational grounds since Chandra and XMM-Newton observations have revealed a significant number of XBONG candidates (Fiore et al. 2000; Comastri et al. 2002; Georgantopoulos & Georgakakis 2005; Civano et al. 2016). However, the nature of XBONGs still needs to be understood.

One possibility is that they could be intrinsically luminous but heavily obscured AGNs, where the covering factor of the obscuration is significant enough that neither broad nor narrow lines escape. This scenario is interesting because they may be part of a missing population of hard X-ray sources necessary to explain the observed X-ray background emission (Gilli, Comastri & Hasinger 2007; Ueda et al. 2014). Another possible explanation is the dilution of nuclear emission lines by the bright starlight of the host galaxy (Moran et al. 2002), i.e., type 2 AGNs with stellar light bright enough to outshine the AGN signature. These galaxies are often called optically dull AGNs (OD AGNs).

Alternatively, XBONGs could be groups of galaxies with a large amount of intragroup medium (IGM) compared to the typical interstellar medium (ISM) of a single galaxy. These groups may not be recognized in typical optical observations if they are poor or even fossil groups, the end products of galaxy mergers dominated by a single elliptical galaxy (e.g., Ponman et al. 1994; Jones et al. 2003; also called an X-ray over-luminous elliptical galaxy – OLEG- by Vikhlinin et al. 1999 and isolated OLEG (IOLEG) by Yoshioka et al. 2004).

Using the large CGC, we can explore XBONG candidates. The X-ray spectral shape (either by hardness ratios or absorbing $N_H$ columns) is one of the critical quantities which can separate the obscured AGNs (hard) and hot gas (soft). The spatial extents can further help to identify hot gas and the temporal variations to identify AGNs. Additionally, WISE IR colors can help to assess the amount of obscuration. Investigating the X-ray spectral, temporal, and spatial characteristics and WISE colors of the XBONG candidates, we will address the origin and nature of the unknown population (Kim et al. 2023 in prep).

**Appendix A. NWAY false match rate**

We have run extensive simulations to determine the rate of false matches and to optimize the match statistics. The simulation is done the same way as in the real match (Section 4), ensemble by ensemble, but after shifting the source positions in eight directions (horizontal, vertical, and diagonal) by 30 arcseconds. Figure 2 shows the matched sources from each catalog in galactic coordinates on the Aitoff projection. The number of total matches ranges from 40K to 110K. The source density varies considerably across the sky.

Since the rate of false positives is a strong function of source density, we performed the simulations across a wide range of X-ray and optical/IR source densities. Figure A1 illustrates the distribution of the total number and density of sources in each ensemble from the catalogs. Note that CSC2 consists of 4380 ensembles, each covering a small part of the contiguous sky. The number of sources per ensemble ranges from 1 to ~ 500,000, and the source density from 0.0003 to ~ 300 sources per arcmin$^2$. The source number and density of the data sets used in the simulations are also marked (color coded) at the top of the two panels of Figure A1.

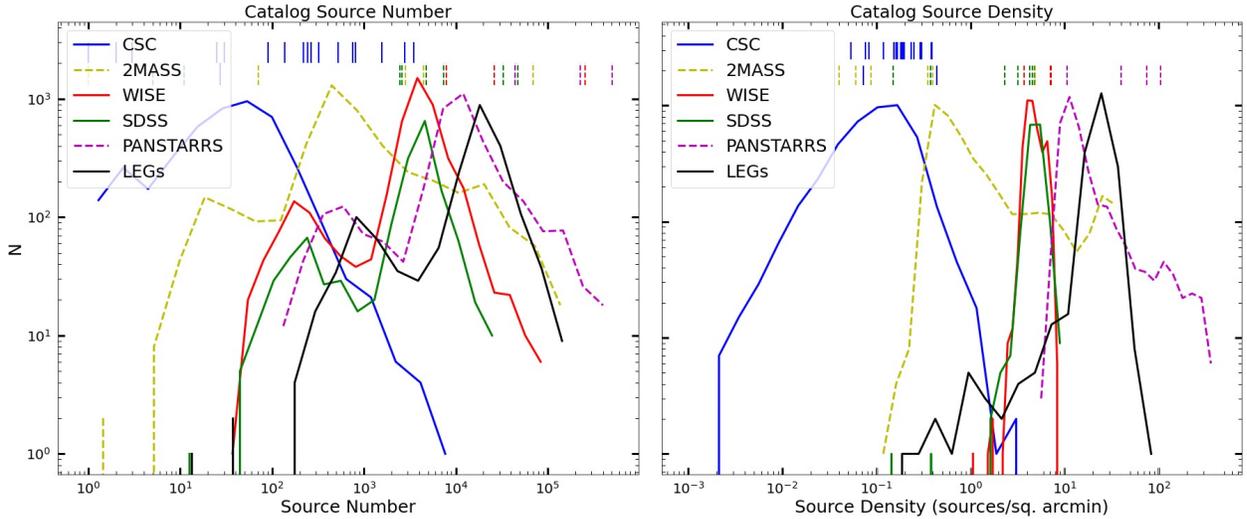

Fig. A1 The total number and density (in arcmin$^2$) of X-ray/optical/IR sources in each ensemble. Unique colors and line styles mark different catalogs (see legend). The ensembles used in the simulation are marked by the color and line style of the catalogs by the small vertical bars at the top of the figure.

Figure A2 shows the estimated false positive rates as a function of optical/IR source density. The rate of false positives is determined by dividing the average number of matches found with the simulation for a given ensemble ("fake matches") by the number of matches identified with the original source positions ("real matches"). Because the X-ray source density is always small (< 3 in arcmin$^2$), the false match rate primarily depends on the optical/IR source density. The top-left panel of Figure A2 indicates the false positive rate when no additional selection in separation is applied, i.e., with all counterparts with separation < 10″ set by the initial parameter. When the source density is low (< 10 sources in arcmin$^2$), the rate of false positives is also low (< 10%). When the source density is intermediate (10 – 30 sources in arcmin$^2$), the rate of false positives is about ~20%. When the source density is very high (> 40 sources in arcmin$^2$), the rate of false positives is also very high (~50%).

The top-right panel of Figure A2 shows the rates after applying separation < 3″. The false positive rate is significantly reduced (almost by half). The rate is always lower than 5% when the source density is less than 10. The horizontal line in the figure indicates the 5% level. When the source density is intermediate (10-30), the rate of false positives is about ~10%. When the source density is very high (>40), the false rate is still as high as 40%. Based on the simulation results, we choose (1) to apply the final cut at separation < 3″ and (2) to exclude the regions of very high source densities (see below).

For the 2$^{nd}$ purpose, we remove the Galactic plane (|b| > 15), entire ensembles with high optical/IR source density, such as nearby large galaxies, M31 and M33, and parts of ensembles within a given radius from known objects, such as Galactic globular clusters (e.g., M2, M3, M5). In our matches between CSC2 and optical/IR catalogs, we achieve a rate of the false positive match of ~5% after excluding these crowded fields and applying our strict selection criteria.

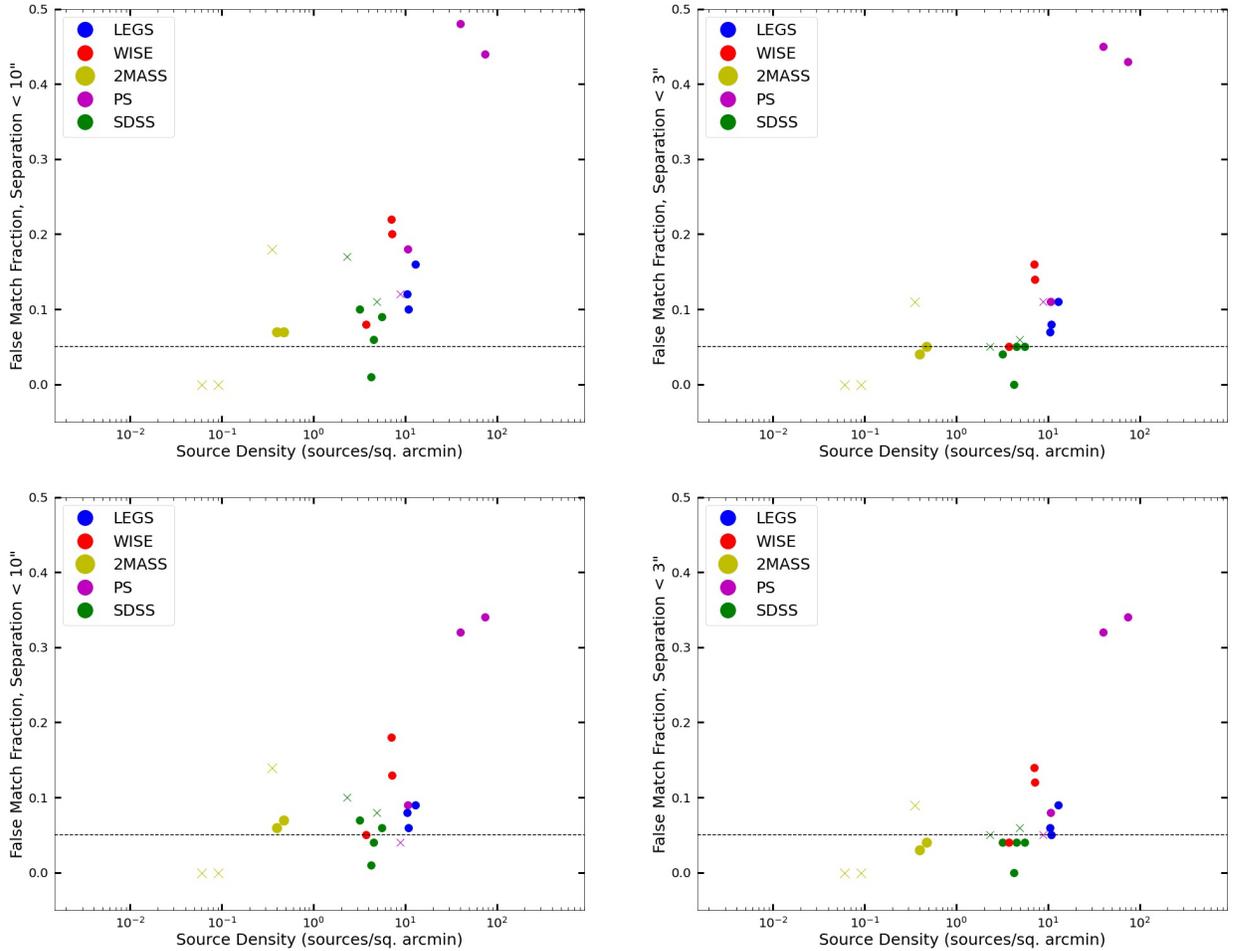

Fig. A2. False match fractions as a function of optical/IR source density. (Top left) Applying the initial maximum separation of 10″ with no further selection by separation, and (top right) Applying an additional selection with separation < 3″. In both cases, we set p_any > 0.5. (Bottom panel) same as the top panel, but with p_any > 0.8. The crosses indicate the ensembles with a relatively small number of counterparts (≲ 10), which are less significant. The horizontal lines indicate that the false positive rate is 5%, a typical limit in this work.

To select the optimal NWAY parameter, p_any (the probability that one of the associations is correct), we test how the false match rate decreases with increasing p_any. Increasing p_any from p_any > 0.5 to p_any > 0.8 (the bottom panel of Figure A2), we find that the number of false matches decreases, but the number of real matches decreases, too, in a way that the reduction of the false positive rate is only minimal. Therefore, we choose p_any > 0.5 to select the final matches.

**Appendix B. Galaxy Zoo Morphological Types**

We extracted morphological types from the Galaxy Zoo 1 release in the SDSS zooVotes table (Linton et al. 2008). In the *spec-z* and *photo-z* samples, we find 1249 objects with listed morphological types. The zooVotes table contains the fraction of votes for six categories: ellipticals, spirals (clockwise, counter-clockwise, and edge-on), mergers, and stars/don't know. We classified objects with a fraction of votes > 0.5 as belonging to that category. We note that the Zoo classifications with vote fractions below 0.7 may be less reliable. The number of sources that fall into each morphological type is listed in Table B1. About three-quarters are elliptical galaxies. This X-ray and Zoo selected sample is somewhat biased against less massive spiral galaxies.

```
Table B1. SDSS Morphological Types
─────────────────────────────────

Type                          #
─────────────────────────────────

Ellipticals (p_el>0.5)       805
Spirals     (p_cs>0.5)       226
Mergers     (p_mg>0.5)        38
Don't Knows (p_dk>0.5)        17
─────────────────────────────────
```

**Appendix C. Comparison of Redshifts**

The spectroscopic redshifts (spec-z) and photometric redshifts (photo-z) for the SDSS counterparts of CSC2 sources are taken from SDSS DR16 SpecObj and Photoz tables: 2,959 objects have both spec-z and photo-z available. In Figure C1, photo-z is plotted against spec-z. The dashed lines indicate the difference of 0.2 dex, i.e., | log(photo-z) -log(spec-z) | = 0.2. Different types are color-coded. Overall, photo-z follows spec-z well in the mid-range of z = 0.03 - 0.5. However, the photo-z deviates from spec-z at lower z (< 0.03; primarily for galaxies), where photo-z is often greater than spec-z. At higher z (> 0.5, primarily for QSOs), photo-z is often lower than spec-z. In Table C1, we list the total number of sources and the fraction of outliers beyond 0.2 dex. Compared with the SDSS spec-z of galaxies, 7% of the SDSS photo-z are inconsistent. This outlier fraction is considerably higher for QSOs (21%).

In the right panel of Figure C1, we select only those with high-quality measurements. First, we apply zWarning=0 for spec-z and photoErrorClass=1 for photo-z (see Beck et al. 2016). This excludes ~15% in this sample with both spec-z and photo-z. Among the objects with only photo-z, the fraction is significantly higher to ~50%. Nonetheless, we apply this strict selection rule. Then, we apply error(z)/z < 0.5 to further remove objects with significant uncertainties. This condition removes ~1% of the sample. As in Table C1, the outlier fraction is reduced to 5% for galaxies and 11% for QSOs. Note that this 5% is comparable to the similar fraction of the false match rate already given when cross-matching CSC2 and other catalogs. While we try to reduce these uncertainties, we consider them an inevitable limit.

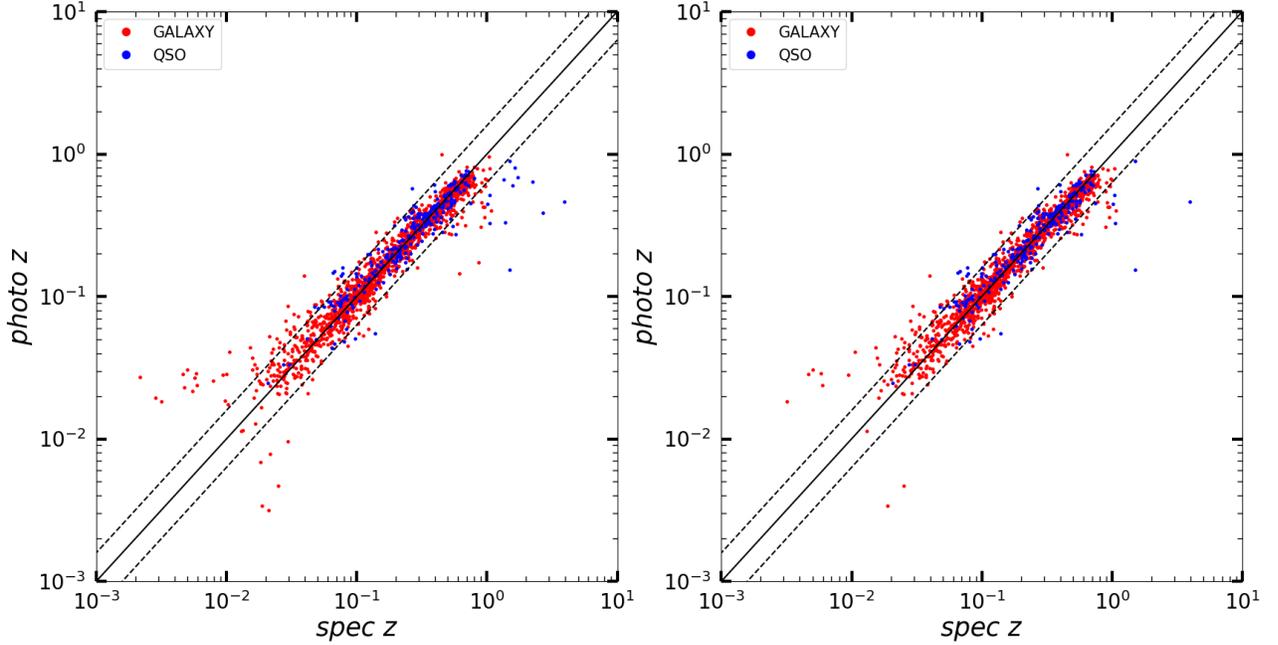

Fig C1. (left) Comparison of SDSS spec-z and photo-z. (right) Poor quality measurements are excluded by selecting only those with zWarning=0 (for spec-z), photoErrorClass=1 (for photo-z), and z / error(z) > 2 (for all).

```
Table C1.  Fraction of outliers when compared with SDSS spec-z
______________________________________________________________________

                 SDSS photo-z       Pan-STARRS photo-z      Legacy photo-z
______________________________________________________________________

            total    outlier      total    outlier       total    outlier

Galaxies    2141   144 (  7%)     1649   176 ( 11%)       917    42 (  5%)
QSOs         813   171 ( 21%)      339   104 ( 31%)       286    48 ( 17%)
Stars          5     5 (100%)        3     3 (100%)         2     2 (100%)

after screening for quality control*
Galaxies    1926    87 (  5%)     1550   134 (  9%)       915    42 (  5%)
QSOs         619    67 ( 11%)      280    67 ( 24%)       282    47 ( 17%)
Stars          2     2 (100%)        0     0               1     1 (100%)
______________________________________________________________________
```

Table C2. Fraction of outliers when compared with SDSS photo-z

|  | Pan-STARRS photo-z | | Legacy photo-z | |
|---|---|---|---|---|
|  | total | outlier | total | outlier |
|  | 3217 | 178 ( 6%) | 2567 | 183 ( 7%) |
| after screening for quality control* | 2980 | 141 ( 5%) | 2551 | 176 ( 7%) |

\* screening For quality control
  zWarning=0 for SDSS spec-z
  photoErrorClass=1 for SDSS photo-z
  extrapolationClass=0 and extrapolationPhotoz=0 for Pan-STARRS photo-z

The photo-z data for Pan-STARRS and Legacy are from Beck et al. (2021) and Zou et al. (2019), respectively. These data are compared with SDSS spec-z in Figure C2. In addition to the screening described above, we only select those with no extrapolation in Pan-STARRS photo-z (by extrapolation_Class=0 and extrapolation_Photoz=0; see Beck et al. 2021). The two photo-z data are consistent with the SDSS spec-z (as good as the SDSS photo-z). The galaxy outlier fractions are 9% and 5% for Pan-STARRS and Legacy, respectively. Again, the outlier fraction for QSOs is higher (17-24%).

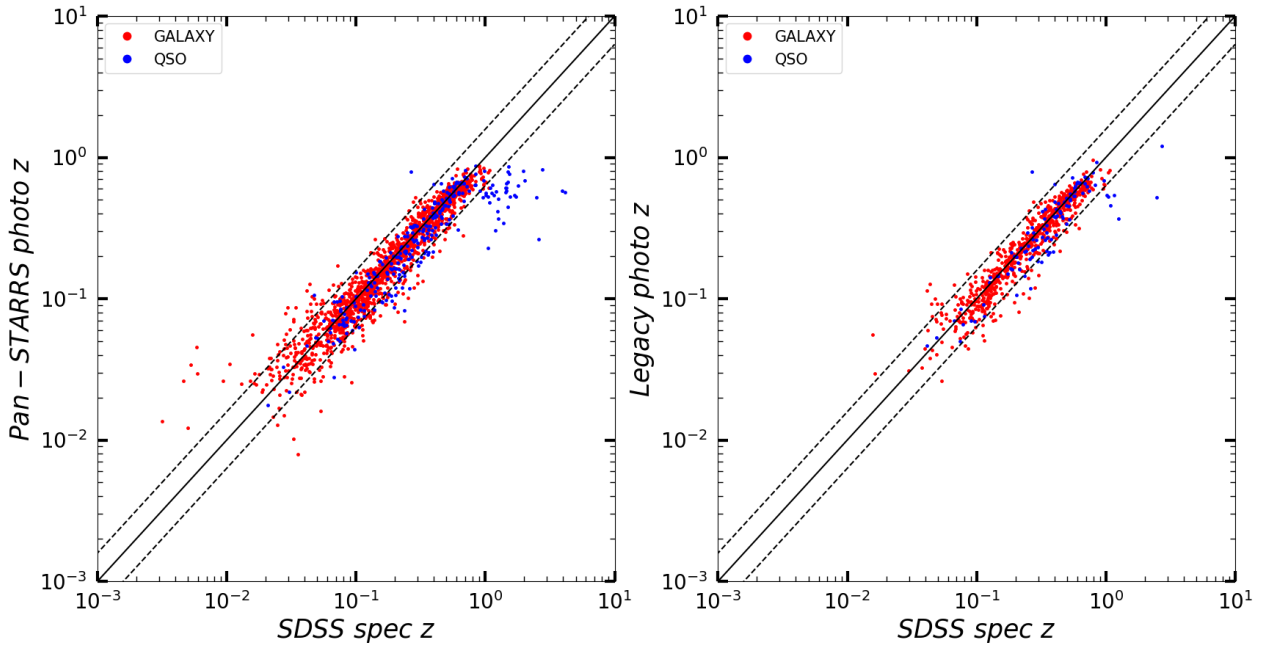

Fig C2. (left) Pan-STARRS photo-z and (right) Legacy photo-z are compared with SDSS spec-z. Poor quality measurements are excluded by selecting only those with zWarning=0 (for SDSS spec-z), extrapolationClass=extrapolationPhotoz=0 (for Pan-STARRS photo-z), and error(z)/z <0.5 (for all).

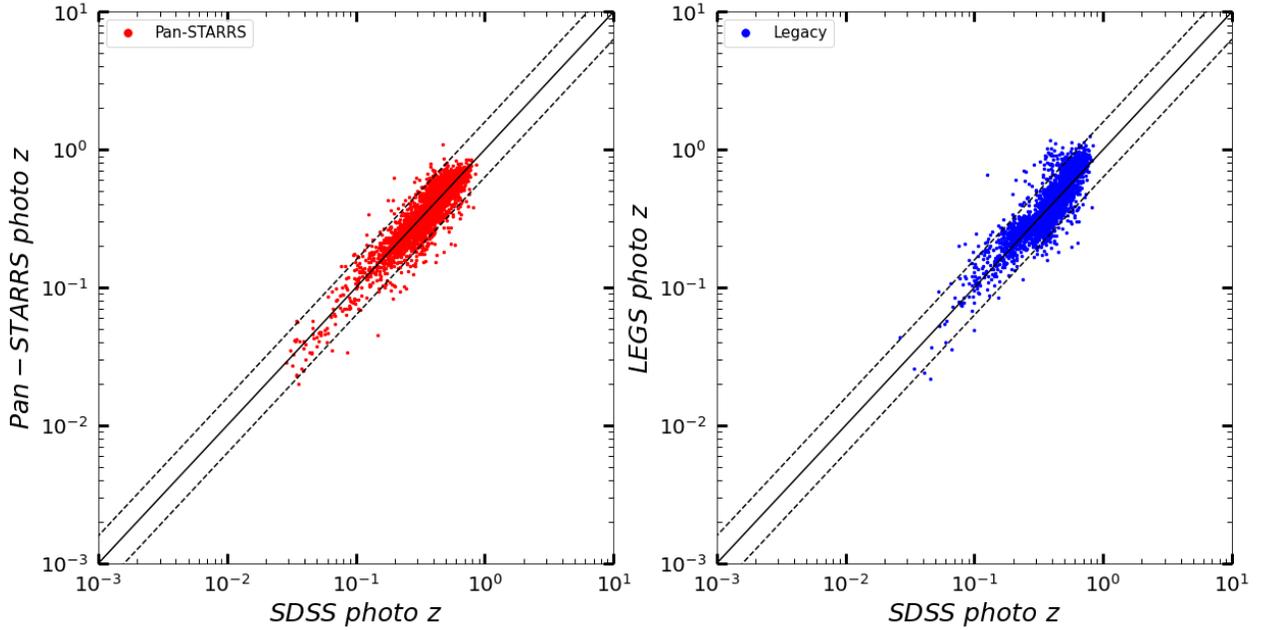

Fig C3. (left) Pan-STARRS photo-z and (right) Legacy photo-z are compared with SDSS photo-z. Poor quality measurements are excluded by selecting only those with photoErrorClass=1 (for SDSS photo-z), extrapolationClass=extrapolationPhotoz=0 (for Pan-STARRS photo-z), and error(z)/z < 0.5 (for all).

We also compare the two photo-z data with SDSS photo-z in Figure C3. No information about their classes is available for these samples, except the expectation that they primarily consist of galaxies. The photo-z data are consistent with the SDSS photo-z. The outlier fractions are 5% and 7% for Pan-STARRS and Legacy, respectively (Table C2).

**Appendix D. Comparison of Optical Magnitudes**

We compare the optical r mag from SDSS, Pan-STARRS, and Legacy catalogs in Figure D1. We take r from the SDSS PhotoObj table[16]. This is the model mag to consider the extended source emission beyond the PSF properly. Similarly, we take rMeanKronMag from Pan-STARRS MeanObject table (Flewelling et al. 2020) and mag_r from Legacy ls_dr8.tractor_s table[17]. The mean difference is 0.15 ± 0.4 (0.2 ± 0.2) between SDSS and Pan-STARRS (Legacy). Note that the scatter is small at the bright end (r < 19 mag)) but significantly increases at the faint end (r > 19 mag). Overall, 90% of them are consistent within 0.5 mag.

---

[16] https://skyserver.sdss.org/dr16/en/help/browser/browser.aspx#&&history=description+PhotoObjAll+U
[17] https://www.legacysurvey.org/dr8/description/

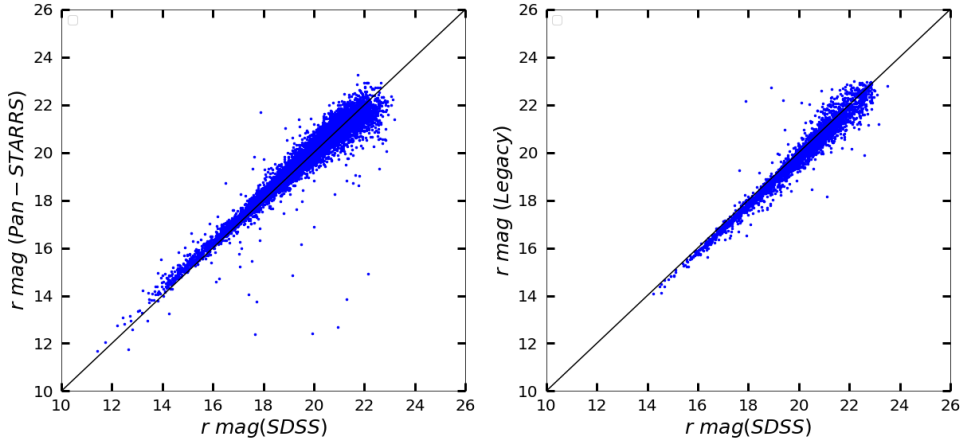

Fig D1. (left) The Pan-STARRS r mag and (right) the Legacy r mag are compared with the SDSS r mag.

**Appendix E. Local Galaxies Exclusion**

Our matching procedure assumes that each galaxy corresponds to one CSC2 source. If multiple CSC2 sources belong to one galaxy, our cross-matching procedure would have listed the central source and underestimated the total X-ray flux. Unlike distant galaxies in typical deep surveys (e.g., Chandra Deep Field, COSMOS), local galaxies, which may have multiple X-ray sources, need special attention.

First, we have excluded all the CSC2 sources with extent >30 arcsec (see Section 2), i.e., most nearby galaxies within ~100 Mpc or $z < 0.025$, where extended and complex X-ray emission is detected with Chandra. These nearby galaxies contain multiple X-ray sources, including diffuse hot gas, X-ray binaries (HMXBs and LMXBs), and possibly low-luminosity AGN. We refer to the Chandra Galaxy Atlas (Kim et al. 2019) and XMM-Newton Galaxy Atlas (Islam et al. 2021) for more information on the nearby galaxies.

Second, for the galaxies within a few x 100 Mpc (roughly $z<0.1$), there may be additional, off-center CSC2 X-ray sources that could be detected but not included in the match, underestimating the total galaxy flux/luminosity. To estimate how many off-center X-ray sources may be missed in our galaxy sample, we have examined the 2MASS redshift survey (2MRS), which includes 44599 local ($z \lesssim 0.1$) galaxies (Huchra et al. 2012) with 98% completeness down to Ks = 11.75 mag.

We find that 417 2MRS galaxies have our selected galaxy candidates within their boundaries (the total mag extrapolation radius in 2MRS). In 128 of these galaxies, the CSC2 source lies at large radii > 3" from the center. These X-ray sources are likely background galaxies as identified in *spec-z* or *photo-z* samples (see section 5). In 160 galaxies we find only one central CSC2 source and in 129 galaxies a central CSC2 source plus an additional off-center CSC2 source. For 19 of the latter, the off-center sources are identified with background galaxies or QSOs in the SDSS *spec-z* sample (see Section 5.1). In the remaining 110 galaxies the off-center sources do not have a crossmatch counterpart. Conservatively, we exclude these 110 galaxies from the CGC because the unidentified off-center sources may belong to the galaxies. We will present local galaxies with multiple X-ray sources in a separate paper.

Comparing the 2MRS 289 galaxies with a central CSC2 source with the SDSS values, we found that 136 of them have redshifts in agreement with SDSS spec-z, within 1%. One exception is IC 2475, where the difference is 30%, for which we adopted the SDSS spec-z value which is consistent with other measurements available in NED. For the remaining 153 2MRS galaxies there is no SDSS spec-z, and the photo-z values (including 131 from PanSTARRS) are different from the 2MRS spec-z. Of these, 66 of them are included in the 110 galaxies that we have excluded from our catalog (see above). For the remaining 87 galaxies, we adopted the 2MRS redshifts.

We noticed that some globular clusters (e.g., from the SLUGGS Survey; Forbes et al. 2017) in nearby galaxies are spectroscopically classified as galaxies in the SDSS SpecObj table. Because of their proximity and low optical luminosity, we exclude 10 nearby ($z < 0.01$), optically faint ($L_r < 10^7\ L_{r\odot}$) objects. All of them are from the SDSS spec-z sample.

## Appendix F. Redshift and $L_X$ Distributions of Galaxies from Different Samples

In this appendix, we test whether different samples (e.g., spec-z vs. photo-z samples) have different global properties (e.g., in redshift and $L_X$), which in turn affect the galaxy finding statistics (as in Tables 5, 8, 10, 11). In Figure F1 (left), we show the z distributions of different samples, including the spec-z sample from SDSS and the photo-z samples from SDSS, PanSTARRS, and Legacy. The z distributions do not vary significantly from one sample to another, particularly at higher z ($z > 0.3$). The only noticeable difference is a slightly higher fraction of galaxies at lower z ($z < 0.1$) in the SDSS spec-z and PanSTARRS samples than in the SDSS photo-z and Legacy-S samples. In Table F1, we list the mean and the standard deviation (std) of the z distributions. The mean values are similar within half of std. We also applied a few different z ranges to examine whether our galaxy selection statistics depend on the different z distributions between the spec-z and photo-z samples. As expected, based on the insignificant difference among different samples, we do not find significantly different statistics.

Similarly, in Figure F1 (right), we show the $\log(L_X)$ distributions of different samples. Again, they are similar, with one noticeable exception - all galaxies with higher $L_X$ ($> 10^{44}$ erg s$^{-1}$) are from the SDSS spec-z sample. Those high $L_X$ galaxies could not be selected in the photo-z samples, because of the significant QSO contaminations. These are the XBONGs discussed in Kim et al. (2023). The mean and std of the $\log(L_X)$ distributions of different samples are listed in Table F1. Again, the mean values are similar within one-third of std.

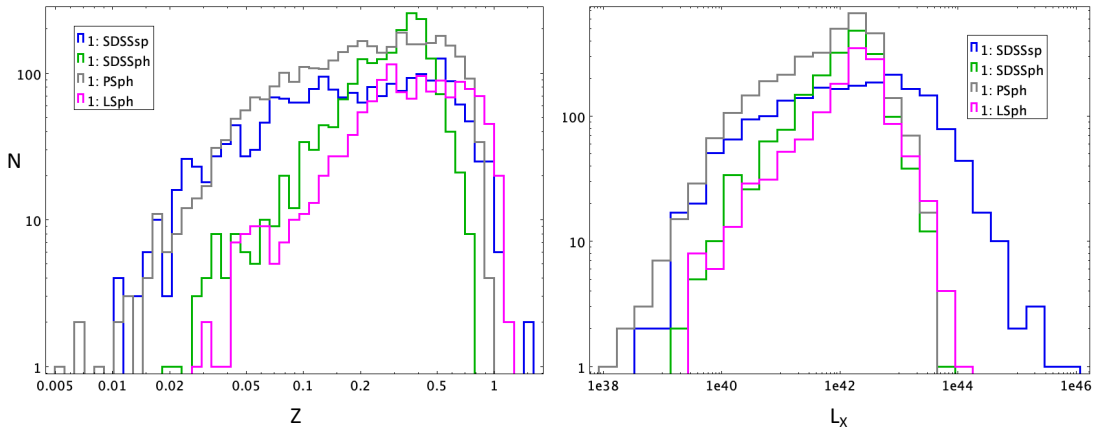

Fig F1. (left) The z distributions and (right) the log($L_X$) distributions of the spec-z sample (SDSS) and the photo-z samples (SDSS, PanSTARRS, Legacy).

Table F1. The z and log($L_X$) distributions of different samples

| | z mean | std | log(Lx) mean | std |
|---|---|---|---|---|
| all | 0.31 | (0.21) | 41.9 | (0.9) |
| SDSS spec-z | 0.27 | (0.23) | 42.1 | (1.2) |
| SDSS photo-z | 0.31 | (0.13) | 42.0 | (0.7) |
| PS photo-z | 0.28 | (0.20) | 41.8 | (0.8) |
| LS photo-z | 0.43 | (0.24) | 42.1 | (0.7) |